\documentclass[12pt]{article}
\usepackage{amssymb,amsmath}
\usepackage{latexsym}
\usepackage{graphicx}
\usepackage{color}
\usepackage{psfrag}
\usepackage[hypertex]{hyperref}

\numberwithin{equation}{section} 

\def\oneone{\rlap 1\mkern4mu{\rm l}}

\def\IR{\mathbb{R}}

\def\cJ{{\cal J}}
\def\cM{{\cal M}}

\def\cQ{{\cal Q}}

\def\cU{{\cal U}}
\def\stratum{superstratum}
\def\strata{superstrata}
\def\nBPS#1{$\frac{1}{#1}$-BPS}

\def\Qt{\widetilde{Q}}
\def\tr{\mathop{\mathrm{tr}}\nolimits}
\def\zh{\widehat{z}}
\def\Ph{\widehat{P}}
\def\thetah{\widehat{\theta}}
\def\Pih{\widehat{\Pi}}

\def\be{\begin{equation}}
\def\ee{\end{equation}}

\begin{document}

\begin{titlepage}

\begin{flushright}
IPhT-T11/151
\end{flushright}

\centerline{\Large \bf  Double, Double Supertube Bubble}
\date{\today}

\bigskip
\centerline{{\bf Iosif Bena$^1$, Jan~de~Boer$^2$,}}
\centerline{{\bf Masaki Shigemori$^{\, 3}$ and Nicholas P. Warner$^4$}}
\bigskip
\centerline{$^1$ Institut de Physique Th\'eorique, }
\centerline{CEA Saclay, F-91191 Gif sur Yvette, France}
\bigskip
\centerline{$^2$Institute for Theoretical Physics, University of Amsterdam,}
\centerline{Science Park 904, Postbus 94485, 1090 GL
Amsterdam, The Netherlands}
\bigskip
\centerline{$^3$Kobayashi-Maskawa Institute for the Origin of Particles and the Universe,}
\centerline{Nagoya University, Nagoya 464-8602, Japan}
\bigskip
\centerline{$^4$ Department of Physics and Astronomy,}
\centerline{University of Southern California,} \centerline{Los
Angeles, CA 90089, USA}
\bigskip
\centerline{{\rm iosif.bena at cea.fr,~J.deBoer at uva.nl, } }
\centerline{{\rm shige at kmi.nagoya-u.ac.jp,~warner at usc.edu} }
\bigskip
\bigskip

\begin{abstract}
\noindent We argue that there exists a new class of completely smooth
\nBPS{8}, three-charge bound state configurations that depend upon
arbitrary functions of \emph{two variables}.  These configurations are
locally \nBPS{2} objects in that if they form an infinite flat sheet
then they preserve $16$ supersymmetries but even with arbitrary
two-dimensional shape modes they still preserve $4$
supersymmetries. They have three electric charges and can be thought of
the result of two successive supertube transitions that involve adding
two independent dipole moments and giving rise to the arbitrary
two-dimensional shape modes.  We further argue that in the D1-D5-P
duality frame this construction will give rise to smooth, horizonless
solutions, or microstate geometries.  We expect these solutions to be
extremely important in the semi-classical and holographic descriptions
of black-hole entropy.
\end{abstract}

\end{titlepage}



\section{Introduction}

One of the most interesting brane configuration to have been discovered
in the last two decades is probably the supertube. In its original
incarnation \cite{Mateos:2001qs}, the supertube is a moving D2 brane in which
D0 branes and F1 strings are dissolved. What makes this D2 brane special
is the fact that the D0 and F1 densities satisfy a relation that allows
the D2 brane to have an arbitrary shape that can follow any closed curve
in the eight dimensions transverse to the F1 and yet remain
supersymmetric \cite{Mateos:2002yf}. The eight Killing spinors preserved by
the supertube are exactly the same as the common Killing spinors of its
``electric'' components: the D0 branes and F1 strings.

The fact that there are \emph{supersymmetric} string theory
configurations determined by arbitrary continuous functions might appear
unexpected, especially if one is used to thinking about the
supersymmetries preserved by infinite, flat branes at
angles. Nevertheless, one can dualize the supertube into a fundamental
string that carries an arbitrary left-moving momentum profile, and
supertubes of various shapes are simply dual to various ways of putting
BPS momentum modes on the fundamental string \cite{Lunin:2001fv,
Dabholkar:1995nc}.

Another interesting feature of supertubes is that, in the duality frame
in which the electric charges are those of D1 and D5 branes, the
back-reacted supertube solution can be made into a smooth geometry
\cite{Emparan:2001ux, Lunin:2001jy, Lunin:2002iz,
Kanitscheider:2007wq}. Moreover, since this solution can be put in an
asymptotically $AdS_3 \times S^3 \times T^4$ geometry, the back-reacted
supertube geometries can be related to various half-BPS microstates of
the D1-D5 CFT, and the entropy of the half-BPS D1-D5 system can be
reproduced by counting the smooth horizonless supertube configurations
\cite{Lunin:2001jy, Palmer:2004gu, Bak:2004kz, Rychkov:2005ji}.
This has led to the
conjecture by Mathur that similar physics will be at work in the
three-charge D1-D5-P system, and therefore the entropy of the D1-D5-P
black hole will come from string and brane configurations that do not
have a horizon and have unitary scattering in the same region of moduli
space in which the classical black hole exists (see \cite{Mathur:2005zp,
Bena:2007kg, Skenderis:2008qn, Balasubramanian:2008da, Chowdhury:2010ct}
for reviews).
When such configurations are smooth, horizonless solutions of
supergravity, they are often referred to as \emph{microstate geometries}
because they represent microstates of the black hole both
semi-classically and through the AdS/CFT correspondence.  It is
reasonable to expect that many of the microstates of the black hole will
be dual to geometries that involve Planck-scale details that go beyond
the validity of the supergravity approximation.  On the other hand, it
is hoped that, within the validity of the supergravity approximation,
one can find a suitably dense, representative sample of microstate
geometries that will not only give a semi-classical picture of the
microstates but also yield some of the thermodynamic details of the full
system and perhaps even reproduce the entropy of the black hole. 
Many such configurations have been
constructed, both in supergravity and using non-back-reacted branes, but
so far the entropy of the back-reacted configurations is not of the same
order as that of black holes with similar charges \cite{deBoer:2009un,Bena:2010gg}. 

One of the common features of the geometries and brane configurations
constructed so far is that they depend either on a finite number of
parameters, or they come from putting arbitrarily-shaped supertubes in
various three-charge geometries and thus depend on several functions of
one variable.  However, two of the authors have recently proposed \cite{deBoer:2010ud} that
there may be BPS string configurations that depend on functions of
\emph{two} variables, and these potentially have much more entropy than
that of the the systems constructed so far.  It is our purpose in this
paper to present evidence that such brane configurations do indeed exist
and that they preserve the same supersymmetries as those of the D1-D5-P
black hole and are determined by several functions of {\em two}
variables.  We will refer to such objects as \emph{\strata}\footnote{For
picture of what is intended here, see the Strata Tower
http://www.dezeen.com/2008/05/13/strata-tower-by-asymptote/ or Corkscrew
Peak http://www.summitpost.org/corkscrew-peak/617471}.

In Section \ref{physdesc} we describe exactly how and why \strata\
can be constructed as smooth, $\frac{1}{8}$-BPS solutions that depend
upon two variables.  In Section \ref{sect:GenBubbling} we summarize how
one can obtain and analyze the supersymmetries in a supertube
transition. We have also included a much more systematic development of
this process in the Appendix.  In Section \ref{sect:DoubleBubbling} we
combine two such supertube transitions to make the ``double bubbled''
\stratum\ and examine its supersymmetry structure and substantiate the
physical description of Section \ref{physdesc}.  We then make some final
remarks in Section \ref{Conclusions}.

\section{The physical description of \strata}
\label{physdesc}

To understand our approach to establishing the existence of \strata\ it is important to recall some of the defining properties of supertubes and the methods by which one can establish the existence of these arbitrary-shaped, supersymmetric configurations.

In string theory it is easy to create many species of two-charge,
\nBPS{4} states by a simple superposition of compatible D-branes,
momentum states or other solitons.  Each charge component is \nBPS{2}
and compatibility means that the \nBPS{2} supersymmetry projectors for
each charge commute with one another so that the two components do not
interact.  The resulting object is not really a new fundamental
object in string theory, and without string corrections and
back-reaction it is really only a marginally bound superposition of the
two components.  The transition to a supertube fuses these two
components into a new \emph{fundamental} bound state and this is achieved by
giving the system an additional dipole moment and angular momentum in a
transverse direction, so that the entire configuration follows a new and
arbitrary profile transverse to the original progenitor configuration.

The resulting object is still a \nBPS{4} state but very close to
the supertube the supersymmetry is locally enhanced to \nBPS{2}. In other words, 
if the supertube profile
were straight and the configuration were a flat sheet then it would be
exactly \nBPS{2} with the preserved supersymmetries depending on the
orientation of the sheet.  For an arbitrary supertube profile the $16$
local supersymmetries depend upon the direction of the tangent to the
profile, but there is always a set of $8$ \emph{common} supersymmetries
that are preserved independent of the profile, and these
supersymmetries are precisely those of the original two-charge system
before the supertube transition.  Thus one of the hallmarks of the
supertube transition that distinguishes it from the progenitor
two-charge superposition is the emergence of this local \nBPS{2}
structure.

Every two-charge system has a supertube transition, and they can all be related by dualities. However, the physics underlying the supertube transition and the way the local \nBPS{2} structure emerges is different in different duality frames.  One of the simplest supertubes has D0 and F1 ``electric'' charges dissolved in a rotating D2-brane \cite{Mateos:2001qs}.  The \nBPS{2}, near-tube limit is simply an infinite, flat D2-brane and the easiest way to understand the emergence of the $16$ supersymmetries in this limit is to consider the M-theory uplift, in which the entire object a boosted \nBPS{2} M2-brane and the D0 and F1 charges correspond to momentum and winding around the eleventh dimension.  As the orientation of the supertube changes, the set of $16$ supersymmetries varies but there is a common subset of $8$ supersymmetries that is preserved, independent of the orientation of the D2-brane.  This subset of eight supersymmetries is precisely the common set of supersymmetries of the electric (F1 and D0) charges of the underlying system.  

There is another very important feature of the supertube transition:  In some duality frames it ``puffs up'' the brane by adding one dimension to the object but in other duality frames it does not.  For example, the original D0-F1 system is intrinsically $(1+1)$-dimensional but the added dipole charge puffs it up to a $(2+1)$-dimensional D2 brane.   Similarly, the D1-D5 system has codimension $4$, but adding the KKM dipole charge ``smears it out'' into a  codimension $3$ object.  On the other hand, from the perspective of M-theory the puffing up of the  D0-F1 system to a D2 brane is simply a matter of tilting and boosting the $(2+1)$-dimensional M2 brane and there is no gain of dimension: In M-theory it is always a configuration of codimension $8$.  This is because the D0 charge is actually a  momentum charge from the eleven-dimensional perspective, and tilting and boosting the brane momentum simply re-orients the surface and the momentum to create the dipole charge and angular momentum.  This seems to be a general pattern: If one of the electric charges is a momentum then the corresponding supertube has the same codimension as the original object but if neither of the  electric charges is a momentum then the supertube transition necessarily adds an extra dimension to the object, puffing it up so that the codimension of the brane configuration decreases by one.   

There  are two basic approaches to establishing the existence of supertubes.    First there are direct methods using either the Dirac-Born-Infeld (DBI) action or using supergravity.  For example, for  D0-F1 supertubes one can use the DBI action of a rotating D2 brane and induce the D0 and F1 charges using world-volume fields, or one can go  to the T-dual of this supertube in which it is a D1 string with a momentum profile.   Alternatively, one can show that there exists a BPS supergravity solution for any supertube profile.  Moreover,  in the D1-D5 duality frame, this solution is smooth.  Both the DBI and the smooth D1-D5 supergravity descriptions yield the 
supersymmetry structure described above. However, the problem with these direct methods is that they are usually difficult to implement because they involve analyzing the totality of the supertube or constructing a complicated supergravity solution.

\begin{figure}
\includegraphics[height=5cm]{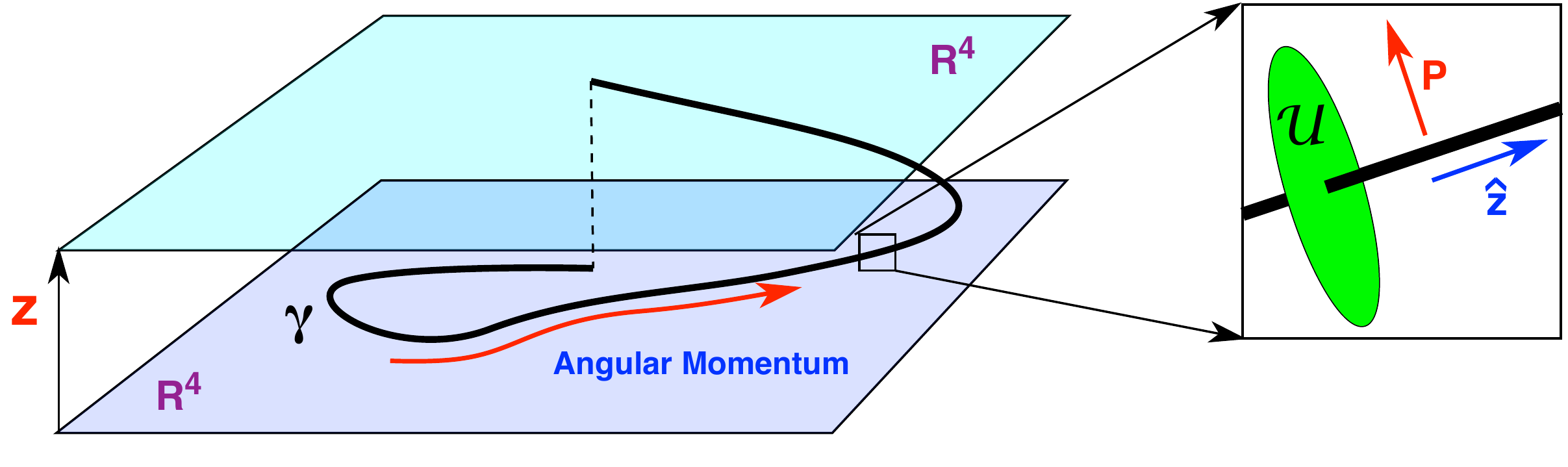}
\caption{The D1-P or  F1-P  supertube profile and a typical local neighbourhood.}\label{zoom}
\end{figure}

There are are also indirect, ``local'' methods that can be used to argue that a certain type of supertube exists.  In this approach one imagines cutting the supertube into very small pieces, or  zooming in very near some point of the tube.  Each such bit of the supertube will look, locally, like an infinite flat brane and will be a \nBPS{2} fundamental object.  However, as the tangent to the supertube profile changes, the set of preserved supersymmetries will also change.  The critical question is then whether there is a subset of  common supersymmetries within all the sets of locally-preserved supersymmetries.   Put differently, the generic situation is that if one takes a brane and tilts it or puffs it up with some dipole charge, then there are no common supersymmetries that are preserved by both the original brane and the tilted or puffed up brane:  The  supersymmetries at different points on the profile are generically completely incompatible.  The remarkable thing that distinguishes the supertube  from the generic brane is that there is indeed a subset of common supersymmetries within the sets of locally-preserved supersymmetries and, by definition, this subset of common supersymmetries is independent of direction of the tangent to the supertube profile.  

The important point is that if one can establish that  each infinitesimal bit of the supertube preserves some set of supersymmetries, and there is a common subset of these supersymmetries that are preserved by all the bits of supertube, then these separate supertube bits are mutually BPS and so do not interact with each other.  
This strongly suggests that they can be strung together to form a supersymmetric, continuous profile of arbitrary shape.   This perspective thus provides a local argument as to why the complicated supergravity configurations can, in fact, exist.

Our purpose in this paper is to apply the supertube process twice in
succession (hence ``double-bubbling'') and then use the supersymmetry
analysis in a local argument to show that, in string theory, there exist
fundamental bound-state configurations of branes that carry three
electric charges, have several dipole moments, preserve $4$
supersymmetries, and are determined by a two-dimensional surface 
arbitrarily-embedded in $\IR^4 \times S^1$.

These $\frac{1}{8}$-BPS configurations, which we refer to as \emph{\strata},  can be thought
of as being made of bits of infinite, flat two-dimensional surfaces,
each bit preserving $16$ supersymmetries, of which $4$ supersymmetries
are common to all the bits, and are the same as the $4$ supersymmetries
common to BPS objects carrying each of the the three electric charges of
the \stratum. As the local argument implies, the fact that all the bits
of the \stratum\ are mutually \nBPS{8} means that the force between
various bits of different orientation will be zero and so one should be
able to assemble them into a \stratum\ whose shape is given by five
arbitrary functions of \emph{two} variables.

The direct approach to finding these \strata\ would be to assemble various types of branes in string or M-theory, find the supersymmetries preserved by these configurations, and vary over all types of branes and all values of the brane densities until one finds a configuration that preserves the same Killing spinors irrespective of two orientations. While this may indeed be  possible, it is  a technically formidable problem.  Our approach, using the doubling of the supertube transition, or double-bubbling,  has several advantages in that it first enables us to identify  what the charges and dipole charges of such a \stratum\  should be, and then determine the conditions these must satisfy in order for the \stratum\ to have an arbitrary shape.  Having achieved this, one then has a good starting point for tackling the far more strenuous and difficult supergravity analysis and solution.  We will defer the latter to a subsequent paper.

Our starting point will be the D1-D5-P system familiar in the three-charge black hole story.  We use this duality frame because it will lead to a smooth configuration.  The first supertube transition will involve adding a dipole moment and angular momentum to take original D1-D5-P system to a three-charge two-dipole charge supertube\footnote{This is dual to a configuration that was originally constructed in the D4-D0-F1 duality frame \cite{Bena:2004wt} as a solution of the Born-Infeld action of a D6 brane of arbitrary shape.  The complete solution has fluxes on the world-volume that induce  D4, D0 and F1 electric charges  and a D2  dipole charge that is related to the other charges.   The corresponding supergravity solution is the same as that of a black ring with only two dipole charges and the relation between the dipole and other charges emerges from the requirement that   the supergravity solution is free of closed timelike curves \cite{Bena:2004wv}.}.  The shape of this generalized supertube is determined by a set of profile functions of one variable. As one approaches the location of the tube, the supergravity solution has a curvature singularity.  If one now zooms in on this three-charge two-dipole charge supertube, or considers the infinite supertube limit, one finds that this infinite tube preserves \emph{eight} supersymmetries, out of which four are the common supersymmetries associated with the component electric charges.  Since the infinite, flat supertube preserves $8$ supersymmetries, it should be a superposition of two mutually-BPS branes and this is most easily seen in the  D1-D5-P  duality frame.

To make the first supertube transition of the D1, D5 and P electric
charges one first partitions the momentum between the D1 and D5 systems to
obtain separate D1-P and D5-P systems. Each of these then undergoes
a supertube transition to objects we will refer to as D1-P and D5-P
supertubes. The details of these transitions will be given later, 
for now it suffices to know that the D1-P and D5-P supertubes
still carry the original charges but acquire angular momentum
and D1 and D5 dipole moments respectively. A cartoon of the D1-P
supertube is given in Fig.~\ref{zoom}.
If $z$ denotes the original common direction of the
D1 and D5 branes, and $\theta$ is the coordinate along the new supertube
profile in $\IR^4$ then the entire configuration now lies along a curve
in the $(z,\theta)$-plane.  We will denote the manifold consisting of
the common direction of the D1 and D5 branes and the four transverse
dimensions by $\cM_5$ and to keep things simple this manifold will be
either  $\IR^4 \times S^1$ or $\IR^5$ depending on whether we compactify
the original brane direction or not.  The three-charge supertube thus has
codimension $4$ and is defined by a curve, $\gamma$, in $\cM_5$.
The D1 (or D5) electric and dipole charges are then simply the $z$ and
$\theta$ components of the total number of D1 (or D5) branes.  This
geometric description immediately implies that the dipole and electric
charges are related by:
\begin{equation}
\frac{Q_1}{ Q_5}   ~=~ \frac{d_1}{ d_5}   \qquad \Leftrightarrow \qquad  Q_1  \,  d_5     ~=~  Q_5  \,  d_1     \,,
\end{equation}
and this is precisely the relation required by either solving the DBI action or by requiring the absence of closed time-like curves in supergravity.   

It is important to remember that in making this supertube transition we have 
fused some of the momentum with the D1 branes and some of the momentum with the
 D5 branes.  The result is parallel D1-P and D5-P supertubes that are each fundamental
 locally-\nBPS{2} objects and together preserve eight supersymmetries locally.  Indeed, the 
generalized ``supertube bit'' is simply a boosted and tilted  superposition
 of D1 and D5 branes.   At this point it also becomes clear why the three-charge
 supertube has a curvature singularity:  this is simply the curvature singularity 
of a solution of superposed D1 and D5 branes in supergravity. 

It is also equally evident how to make a second supertube transition
that fuses the coincident D1-P and D5-P supertubes described above into a new
fundamental object that locally preserves $16$ supersymmetries: One
applies a second supertube transition that involves adding a KKM dipole
charge and angular momentum. Locally, this is the same as the standard
supertube transition of the D1-D5 system.  It is important to remember
that this transition decreases the codimension of the system, and because  
the D1-D5 common direction shrinks smoothly to zero at the KKM profile, the resulting configuration is 
smooth \cite{Lunin:2001jy,Lunin:2002iz}. Hence, the puff-up into a codimension-three object  
\emph{completely resolves the singularity of the D1-D5 system.}

To be more specific, let $\hat z$ denote the common direction of the D1 and
D5 branes before puffing up and recall that there is, locally, a patch,
$\cU$, of $\IR^4$ transverse to the branes (see Fig.~\ref{zoom}).  The
smooth solution is obtained by introducing a KKM dipole charge along a
closed path, $\hat \gamma$, in $\cU$ and smearing the D1 and D5 charge
along this path.  We will parametrize the curve, $\hat \gamma$, by an
angle, $\psi$, so the puffed up brane is a codimension $3$ object that
sweeps out the $(\hat z, \psi)$-plane. The resulting object is now
described by the curve, $\hat \gamma$, in $\cU$ and the
three-dimensional transverse geometry in $\cU$ in the neighborhood of a
point on $\hat \gamma$, appears, at first sight, to be singular.
However, it is a Kaluza-Klein monopole and if the $\hat z$ direction is
compactified with the proper periodicity then the KKM fiber shrinks to zero 
at a certain profile in $\IR^4$ in such a way that the resulting geometry is smooth.

\begin{figure}
\begin{center}
\vspace*{-1cm}
 \hspace*{2cm}\includegraphics[height=12cm]{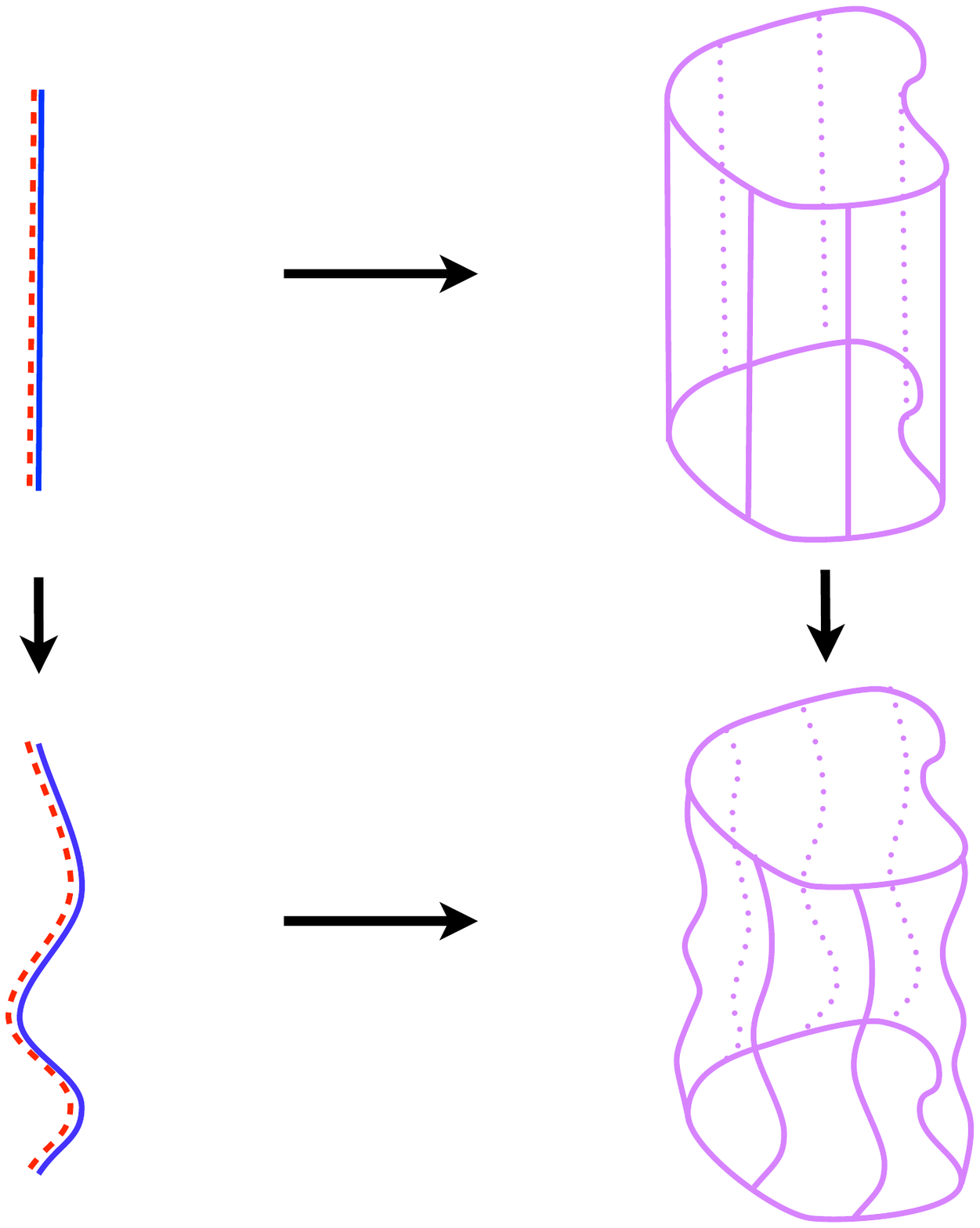}
\vspace*{-1.5cm}
\caption{The
double bubbling of the D1-D5-P system. There are two ways to obtain a
\stratum : The D1 and P can fuse into a D1-P supertube spiral (red
dotted line), and the D5 and P can fuse into a D5-P spiral (blue
continuous line). The spirals can then fuse into a \stratum
. Alternatively the D1-D5 can fuse into a D1-D5-KKM tube (violet
straight supertube), which upon adding momentum can start shaking and
become a \stratum . }\label{superstratum}
\end{center}
\end{figure}

The second supertube transition thus has two very important effects:
First, it completes the fusion of the D1-D5-P system into a true bound
state by fusing the D1-P and D5-P supertubes into a
single, locally \nBPS{2} object.  Secondly, it resolves the singularity
of the generalized three-charge supertube through a KKM puffing
\emph{at every point}, $\hat z$, along the original profile, $ \gamma$,
of the first generalized supertube.  This second supertube transition
puffs the configuration up by one dimension along another arbitrary
curve, $\hat \gamma_{\hat z}$, whose profile can depend upon $\hat z$.
Thus the resolution of the singularity allows the first
(arbitrary-shaped) three-charge supertube profile to be replaced by a
freely choosable curve, $\hat \gamma_{\hat z}$, at every point of the
original profile: In other words, the original three-charge supertube
can be puffed up into a two-dimensional sheet, or \emph{stratum} that has
codimension $3$ in $\cM_5$.  Moreover, since the profile of the second
puff up is freely choosable \emph{at every point} of the original
profile, the resulting sheet, or stratum, is defined by a freely
choosable function, $\vec F(v_1, v_2)$, of two variables into the
$\cM_5$.  This map defines the \stratum\ and it will be a smooth,
$\frac{1}{8}$-BPS configuration that emerges from ``double bubbling.''
The solution is locally \nBPS{2} but globally has the electric charges
of, and the same supersymmetries as, the D1-D5-P system and carries
several dipole charges corresponding to transverse D1 and D5 branes and KK-Monopoles.

The foregoing argument and lays out precisely the construction that leads to the \strata\ and the process  
is depicted in Fig.~\ref{superstratum}.  
However, to substantiate the claim that \strata\ can have a completely arbitrary 
two-dimensional shape 
and still preserve four supersymmetries, we need to complete the ``local argument'' and show
that each locally-flat two-dimensional surface bit that makes a \stratum\ 
preserves $16$ supersymmetries, and that collectively the bits preserve a common 
subset of four supersymmetries.  This will be done in the next two sections.  
 Once this is established, it follows that all the surface bits are
mutually BPS and hence  non-interacting and therefore  they can 
be combined make a complete two-dimensional \stratum.   Of course, to fully establish 
the existence of a \stratum\ we either need to find a Born-Infeld-like description
or to construct its complete supergravity solution. 
We leave this somewhat daunting task for future work.

\section{Supertube transitions}
\label{sect:GenBubbling}

\subsection{Supersymmetries and supertubes in general}

We will consider several examples of supertube transitions and their
effect upon the structure of the supersymmetries.  If $Q_1$, $Q_2$ are
the original electric charges corresponding to some branes and $d_1$ and
$\cJ$ are the dipole charge and angular momentum of the supertube
configuration, then we will use the following canonical notation to
denote the supertube construction:
\begin{equation}
\left( \begin{array}{c}  Q_1 \, (x) \\ Q_2\,  (y)  \end{array}  \right) ~\rightarrow~ \left( \begin{array}{c}  d_1\,   (z\psi)\\  \cJ\,  (\psi) \end{array}  \right)  \,, 
\label{GenSTnotn}
\end{equation}
where $x,y,z$ and $\psi$ are some subsets of coordinate directions along
which the branes are wrapped or the momentum is directed.  The
coordinate, $\psi$, will generically indicate the new direction
associated with the supertube.

Our purpose here is to find the supersymmetries preserved by the supertubes by constructing a one-parameter family of projectors whose null spaces intersect over and define the subspace of supersymmetries that are preserved by the particular supertube configuration.

There is a standard procedure for computing the required supersymmetry projectors:  One starts with the projectors appropriate to the electric charges, $Q_1$ and $Q_2$ and uses carefully selected rotation matrices that tilt and boost the brane configuration along the supertube directions.   The end result is a projector that satisfies three conditions:  (i)  It must be a linear combination of projectors for all the underlying branes and momenta, (ii)  it must define a $\frac{1}{2}$-BPS state (for fixed $x,y,z$ and $\psi$) and so have sixteen null vectors, and (iii) it must be a linear combination of the original projectors associated with the electric branes.   

The first condition simply stipulates that the supertube has the required brane constituents  and the second condition implies that the infinite planar supertube is $\frac{1}{2}$-BPS and preserves $16$ supersymmetries.  If the orientation, $\psi$, of the supertube varies then the set of sixteen supersymmetries also varies, however the last condition guarantees that no matter how the orientation varies the supertube will always preserve the original eight  common supersymmetries associated with the  electric charges and so the supertube can have an arbitrary profile and still be a $\frac{1}{4}$-BPS state.

It turns out that these constraints are  enough to determine the requisite projectors for a supertube configuration and so we will use this approach to derive all the projectors we need.  A more formal and precise derivation of the validity of our projectors can be found in the appendix where we also summarize the dictionary that defines the $\frac{1}{2}$-BPS projectors associated with each and every type of brane charge.

\subsection{A simple example: the F1--P system}
\label{sect:F1P}

To illustrate some basic properties of the supertube transition, we consider
the F1--P system. This system is described in considerably more detail in 
appendix~\ref{sssapp:F1P}. 

The starting point is a certain number of fundamental strings stretched in the $x^1$-direction. These fundamental strings preserve half of the
supersymmetries, namely those that obey 
\begin{equation}
\Pi_{\rm F1}{\cal Q}=0 \,, 
\end{equation}
where 
\begin{equation}
 \Pi_{\rm F1} = \frac{1}{2}(1 + \Gamma^{01}\sigma_3) \,,
\end{equation}
and $\sigma_3$ is the third Pauli matrix which acts on the doublet of Majorana-Weyl supercharges of the type II superstring theory.

To generate the F1-P system, we consider an arbitrary transverse direction,  denoted by $\psi$, and T-dualize the system along a direction in the $(x^1,\psi)$-plane that makes an angle, $\alpha$, with the $x^1$-axis.  Under this T-duality, the system remains \nBPS{2}. It is not very important in this discussion whether the direction along which we T-dualize is compact or not. If it is not compact, what we are doing is not a symmetry of the theory but it can be viewed as a solution-generating operation.

To find the supercharges that are preserved after the T-duality, we (trivially) rewrite the projector in components parallel and perpendicular to the direction of the T-duality axis:
\begin{equation}
\Pi_{\rm F1} = \frac{1}{2}(1 -\sin\alpha
(\cos\alpha \Gamma^{0\psi} -\sin\alpha \Gamma^{01})\sigma_3
+ \cos\alpha (\cos\alpha \Gamma^{01} + \sin\alpha \Gamma^{0\psi})\sigma_3)\,.
\end{equation}
T-duality parallel to fundamental strings converts  them into momentum, and the supersymmetries preserved by momentum are determined by the same projector as  for fundamental
strings but without the $\sigma_3$. Thus the projector we get after the T-duality is simply:
\begin{equation}
\Pi_{\text{F1-P}} = \frac{1}{2}(1 -\sin\alpha
(\cos\alpha \Gamma^{0\psi} -\sin\alpha \Gamma^{01})\sigma_3
+ \cos\alpha (\cos\alpha \Gamma^{01} + \sin\alpha \Gamma^{0\psi}))  
\end{equation}
and this describes the projector for the bound state of momentum along the T-duality axis with some fundamental strings in the orthogonal
direction. By construction, this projector describes a \nBPS{2} system.

If we start with $N$ coincident fundamental strings, the resulting system 
has (F,P)-charges given by $(w,n)=(N\sin^2\alpha,N\cos^2\alpha)$ in the $x^1$-direction,
and (F,P)-charges $(d,J)=(N\sin\alpha\cos\alpha,N\sin\alpha\cos\alpha)$ in
the $\psi$-direction. 

The remarkable feature of this new system is that it preserves a fixed set of eight supercharges regardless of the choice of the direction $\psi$  and the angle $\alpha$. For $\alpha=0$ we have momentum in the $x^1$-direction, and for $\alpha=\pi/2$ we have fundamental strings in the $x^1$ direction, so the eight supercharges are the same as the eight supercharges preserved by parallel fundamental strings and momentum in the $x^1$-direction.  Indeed, we can write 
\begin{equation}
\Pi_{\text{F1-P}} = \sin\alpha (\sin\alpha - \cos \alpha \Gamma^{1\psi})
\Pi_{F1} + \cos\alpha (\cos\alpha + \sin\alpha \Gamma^{1\psi}) \Pi_{P}  \,,
\end{equation}
which clearly demonstrates that the common supersymmetries of a marginal bound state of fundamental strings and momentum in the $x^1$-direction are  always preserved.

For all this to work it is crucial that the amount of F1-string charge and momentum is correlated with the angle $\alpha$. Without this correlation the configuration would have no remaining supersymmetry.

By gluing together pieces of fundamental strings and momentum that locally look like the above F1--P system, we can make a $\frac{1}{4}$-BPS F1--P supertube which is the S-dual of the D1-P supertube shown in Fig.~\ref{zoom}.

\subsection{Bubbling the D1--D5 system}
\label{sect:KKM}

Consider the bubbling
\begin{equation}
\left( \begin{array}{c}  D1 \, (0z) \\ D5\,  (01234z)  \end{array}  \right) ~\rightarrow~ \left( \begin{array}{c}  KKM\,   (01234\psi;z)\\  P\,  (\psi) \end{array}  \right)
  \,. \label{Bubb1}
\end{equation}
In this configuration the special direction of the KKM lies along the common direction, $z$,  of the D1--D5 system and after bubbling the charges and KKM dipole are distributed along the closed curve parametrized by $\psi$.  This is therefore a true ``puffing up'' in that the configuration has gained an extra dimension defined by $\psi$.

Before bubbling, the electric projectors are (see the appendix for details): 
\begin{equation}
  \Pi_{D1} ~=~ {1 \over 2} \, \Big( \oneone ~+~ \Gamma^{0z} \sigma_1 \Big) \,, \qquad     \Pi_{D5} ~=~ {1 \over 2} \, \Big( \oneone ~+~ \Gamma^{01234z} \sigma_1 \Big) 
  \,, \label{Projs1}
\end{equation}
and these two projectors commute. Bubbling combines these and adds a momentum part, $\Gamma^{0\psi}$, and a KKM part, $\Gamma^{01234\psi}$.  

One can then show that the bubbled projector can be written in either of the three following
ways
\begin{eqnarray}
  \Pi  &=&  {1 \over 2} \, \Big( \oneone + \Gamma^{0z}  \big(\cos \beta\,  \sigma_1 \oneone  +  \sin \beta \, \Gamma^{z \psi}  \big) \big(\cos \beta\,   \oneone  -  \sin \beta \, \sigma_1 \Gamma^{1234 z \psi}  \big)  \Big)  \label{BubbProj1a}  \\
  &=& {1 \over 2} \, \Big( \oneone +  \cos^2 \beta \,  \Gamma^{0z} \sigma_1 +  \sin^2 \beta \, \Gamma^{01234z}  \sigma_1   ~+~ \sin  \beta  \cos \beta\, \big(\Gamma^{0\psi}  -  \Gamma^{01234\psi}\big) \Big)  \label{BubbProj1b}  \\
    &=& \cos \beta   \big(\cos \beta\,   \oneone  -  \sin \beta \, \sigma_1 \Gamma^{z \psi}  \big)  \,   \Pi_{D1}  ~+~  \sin \beta   \big(\sin \beta\,   \oneone  +   \cos \beta \, \sigma_1 \Gamma^{z \psi}  \big)  \,   \Pi_{D5}  \,.  \label{BubbProj1c} 
\end{eqnarray}
The first expression, (\ref{BubbProj1a}), shows the underlying rotations
and middle expression, (\ref{BubbProj1b}), shows that the projector is,
indeed, a combination of the projectors for component branes. The second
term in (\ref{BubbProj1a}) squares to $\oneone$ and is traceless, and
hence $ \Pi$ preserves sixteen supersymmetries.  The third expression,
(\ref{BubbProj1c}), shows that these sixteen supersymmetries include the
eight supersymmetries in the common nullspace of $\Pi_{D1}$ and
$\Pi_{D5}$. Note that these expressions are very similar to the expressions
obtained for the F1--P system described above, as they should be because the two systems are dual to one another. 

Thus this projector has a sixteen-dimensional null space that depends
upon the orientation of the supertube through the appearance of
$\Gamma^\psi$.  The projectors, $\Pi_{D1}$ and $\Pi_{D5}$, are
independent of $\Gamma^\psi$ and so their eight-dimensional common null
space is independent of the supertube orientation and shape.  As a
result, if the supertube is an infinite flat sheet then $\psi$ has a
constant orientation and it is $\frac{1}{2}$-BPS but if the supertube
has a varying orientation, or shape, then it is still a
$\frac{1}{4}$-BPS configuration.

\subsection{Bubbling the D1--P and D5--P system}
\label{sect:D1D5}

In the D1--P and D5--P systems, one of the electric charges is a
momentum and so the bubbling to a supertube does not ``puff up'' the
supertube because the final supertube configuration has the same
dimension as the original electric configuration. The supertube is rather a ``superhelix,''  and the tilt and boost of the 
electric charges along a transverse direction, $\theta$, convert some of the D-brane charge into dipole charge and some of the
momentum into angular momentum.  Thus we have:
\begin{equation}
\left( \begin{array}{c}  D1 \, (0z) \\ P \,  (z)  \end{array}  \right) ~\rightarrow~ \left( \begin{array}{c}  d1\,   (0 \theta)\\  J  \,  (\theta) \end{array}  \right) \,, \qquad
\left( \begin{array}{c}  D5 \, (01234z) \\ P \,  (z)  \end{array}  \right) ~\rightarrow~ \left( \begin{array}{c}  d5\,   (01234 \theta)\\  J  \,  (\theta) \end{array}  \right)
  \,. \label{Bubb2}
\end{equation}
Indeed, there is an underlying  tilt angle, $\alpha$, that determines how the charges are realigned after tilting:
\begin{align}
Q_1 &= Q_{D1, z} ~=~  Q_{D1} \cos \alpha \,,  & d_1 &=~ Q_{D1, \theta} ~=~  Q_{D1} \sin \alpha \,,    \label{3chga} \\ 
Q_5 &= Q_{D5, z} ~=~  Q_{D5} \cos \alpha \,,  & d_5  &=~ Q_{D5, \theta} ~=~  Q_{D5} \sin \alpha \,,    \label{3chgb} \\ 
Q_P &= Q_{P, z} ~=~  P \sin \alpha \,,   &  J_\theta  &=~ P_{ \theta} ~=~  P  \cos \alpha \,.   \label{3chgc}
\end{align}
Note that (\ref{3chgc}) differs from the first two equations essentially because the momentum is perpendicular to the branes
just as we had in the F1--P system. Notice that $P_{\theta}$ here should be thought of as the momentum in the 
negative $\theta$-direction.

The fundamental projectors associated with this system are:
\begin{equation}
  \Pi_{D1} ~=~ {1 \over 2} \, \Big( \oneone ~+~ \Gamma^{0z} \sigma_1 \Big) \,, \qquad     \Pi_{D5} ~=~ {1 \over 2} \, \Big( \oneone ~+~ \Gamma^{01234z} \sigma_1 \Big) 
  \,, \qquad  \Pi_{P} ~=~ {1 \over 2} \, \Bigl( \oneone ~+~ \Gamma^{0z}\Bigr) \,. \label{Projs2}
\end{equation}
The projectors associated with these two supertube transitions are
\begin{eqnarray}
  \widehat \Pi_{D1}  &=&  {1 \over 2} \, \Big( \oneone + \Gamma^{0z}  \big(\cos \alpha\,  \sigma_1 \oneone  +  \sin \alpha \, \Gamma^{z \theta}  \big) \big(\cos \alpha\,   \oneone  -  \sin \alpha \,  \Gamma^{z \theta}  \big)  \Big)  \label{BubbProj2a}  \\
  &=& {1 \over 2} \, \Big( \oneone +  \cos^2 \alpha \,  \Gamma^{0z} \sigma_1 +  \sin^2 \alpha \, \Gamma^{0z}    ~+~ \sin  \alpha  \cos \alpha\, \Gamma^{0\theta}  \, \big(\oneone - \sigma_1\big) \Big)  \label{BubbProj2b}  \\
    &=& \cos \alpha   \big(\cos \alpha\,   \oneone  +  \sin \alpha \, \sigma_1 \Gamma^{z \theta}  \big)  \,   \Pi_{D1}  ~+~  \sin \alpha   \big(\sin \alpha\,   \oneone  -   \cos \alpha \, \sigma_1 \Gamma^{z \theta}  \big)  \,   \Pi_{P}  \,, \label{BubbProj2c} 
\end{eqnarray}
and
\begin{eqnarray}
  \widehat \Pi_{D5}  &=&  {1 \over 2} \, \Big( \oneone + \Gamma^{0z}  \big(\cos \alpha\,  \Gamma^{1234} \, \sigma_1    +  \sin \alpha \, \Gamma^{z \theta}  \big) \big(\cos \alpha\,   \oneone  -  \sin \alpha \,  \Gamma^{z \theta}  \big)  \Big)  \label{BubbProj3a}  \\
  &=& {1 \over 2} \, \Big( \oneone +  \cos^2 \alpha \,  \Gamma^{01234z} \sigma_1 +  \sin^2 \alpha \, \Gamma^{0z}    ~+~ \sin  \alpha  \cos \alpha\, \Gamma^{0\theta}  \, \big(\oneone -  \Gamma^{1234} \,\sigma_1\big) \Big)  \label{BubbProj3b}  \\
    &=& \cos \alpha   \big(\cos \alpha\,   \oneone  +  \sin \alpha \, \sigma_1 \Gamma^{z \theta}  \big)  \,   \Pi_{D5}  ~+~  \sin \alpha   \big(\sin \alpha\,   \oneone  -   \cos \alpha \, \sigma_1 \Gamma^{z \theta}  \big)  \,   \Pi_{P}  \,, \label{BubbProj3c} 
\end{eqnarray}
where $\Pi_{D1}$, $\Pi_{D5}$ and $\Pi_P$ are given in  (\ref{3chga}), (\ref{3chgb})  and (\ref{3chgc}).

In both of these equations, the middle expressions show that the
projectors are a combination of the appropriate component parts.  The
second term in each of (\ref{BubbProj2a}) and (\ref{BubbProj3a}) squares
to $\oneone$ and is traceless, and hence each projector preserves
sixteen supersymmetries.  The expressions (\ref{BubbProj2c}) and
(\ref{BubbProj3c}) show that each of these new projectors can be
expressed in terms of the original projectors of the D1--P system or
D5--P systems respectively.  The projectors $\Pi_{D1}$, $\Pi_{D5}$ and
$\Pi_{P}$ all commute with one another and are independent of
$\Gamma^\theta$ and so their nullspaces are independent of the supertube
orientation and shape.  As a result, if the supertube is an infinite
flat sheet then $\theta$ has a constant orientation and the supertube is
$\frac{1}{2}$-BPS but if the supertube has a varying orientation, or
shape, then it is still a $\frac{1}{4}$-BPS configuration.

\section{Double Bubbling }
\label{sect:DoubleBubbling}

\subsection{The transition to the three-charge supertube}
\label{sect:3chgtrans}

We now consider the  D1--D5--P system and consider  it to be  a superposition of the D1--P and the D5--P systems considered above with the momentum partitioned  into two parallel  parts, $P = P^{(1)} + P^{(2)}$ associated with the two different sets of branes.  The first supertube is then obtained by tilting and boosting both sets of branes in exactly the same manner:
\begin{equation}
\left( \begin{array}{c}  D1 \, (0z) \\ P^{(1)}\,  (z)  \end{array}  \right) ~\rightarrow~ \left( \begin{array}{c}  d1\,   (0 \theta)\\  J^{(1)} \,  (\theta) \end{array}  \right) \,, \qquad
\left( \begin{array}{c}  D5 \, (01234z) \\ P^{(2)}\,  (z)  \end{array}  \right) ~\rightarrow~ \left( \begin{array}{c}  d5\,   (01234 \theta)\\  J^{(2)} \,  (\theta) \end{array}  \right)
  \,. \label{combinedBubb}
\end{equation}
The charges of the system are then given by  (\ref{3chga})--(\ref{3chgc}) 
and the angular momentum is similarly decomposable into two parts, $J = J^{(1)} + J^{(2)}$.  

Locally, the picture of these supertubes is one where D1 and D5 branes are
tilted and have some momentum in the transverse direction. The momentum
has to be distributed in such a way that the D1 and D5-branes remain
parallel, so locally the split in momenta between the D1 and D5-branes
is not arbitrary but determined by the ratio of the tensions of the D1 and
wrapped D5-branes.

This produces a three-charge, two-dipole-charge supertube that follows
the trajectory defined by $\theta$.  It preserves the supersymmetries
defined by the projectors $\widehat \Pi_{D1}$ and $\widehat \Pi_{D5}$
defined in (\ref{BubbProj2a})--(\ref{BubbProj3c}).  Note that, unlike
the projectors in \eqref{Projs1}, $\Pih_{D1}$ and $\Pih_{D5}$ do not
commute. However their commutator is proportional to
$\Pih_{D1}-\Pih_{D5}$ and so they commute on their common null space and
thus define eight compatible supersymmetries\footnote{In section \ref{apsc:doubleBubbling}
of the Appendix 
we construct a commuting set of projectors for this three-charge,
two-dipole-charge supertube configuration and find that using these
instead of $\Pih_{D1}$ and $\Pih_{D5}$ leads to the same final result.}.  However, these expressions depend upon $\Gamma^\theta$ and so
the eight-dimensional common nullspace depends upon the supertube
orientation.  On the other hand, (\ref{BubbProj2c}) and
(\ref{BubbProj3c}) show that these projectors can be expressed in terms
of the projectors of the D1--D5--P system, (\ref{Projs2}), and the
common nullspace of $\widehat \Pi_{D1}$ and $\widehat \Pi_{D5}$ includes
the four supersymmetries of the D1--D5--P system that lie in the common
nullspace of the projectors (\ref{Projs2}).  Thus the generic
configuration is still $\frac{1}{8}$-BPS.

\subsection{A basis change}

It is convenient to define new gamma matrices:
\begin{equation}
 \Gamma^{\hat z}  ~=~   \cos \alpha \, \Gamma^z - \sin \alpha \,\Gamma^\theta \,,  \qquad  \Gamma^{\hat \theta}  ~=~   \sin \alpha\, \Gamma^z + \cos \alpha\, \Gamma^\theta  \,. 
 \label{newGam}
\end{equation}
In this new basis one has:
\begin{eqnarray}
  \widehat \Pi_{D1}  &=&  {1 \over 2} \, \Big( \oneone + \cos \alpha\,  \Gamma^{0\hat z} \,  \sigma_1   +  \sin \alpha \, \Gamma^{0 \hat \theta}   \Big)  \label{BubbProj4a} \,  \\
    \widehat \Pi_{D5}  &=&  {1 \over 2} \, \Big( \oneone + \cos \alpha\,  \Gamma^{01234\hat z} \,  \sigma_1   +  \sin \alpha \, \Gamma^{0  \hat \theta}   \Big)  \label{BubbProj4b}\,.
\end{eqnarray}
This shows that the projectors of the three-charge, two dipole charge supertube are simply a combination of the fundamental brane projectors along the $(\hat z, \hat \theta)$ directions.

\subsection{The double-bubbled \stratum}

The goal is now to combine the transition in Section \ref{sect:KKM} with
that described in Sections \ref{sect:D1D5} and \ref{sect:3chgtrans}.  The ``quick and dirty'' way  to
achieve this is to essentially replace $\Pi_{D1}$ and $\Pi_{D5}$ in
(\ref{BubbProj1c}) with $\widehat \Pi_{D1}$ and $\widehat \Pi_{D5}$. Since the fundamental branes
associated with three-charge, two-dipole-charge supertube are tilted and lie
along the $\hat z$-direction, one must also replace the $\Gamma^z$'s in
(\ref{BubbProj1c}) with $\Gamma^{\hat z}$'s.  The resulting candidate
projector is:
\begin{equation}
\widehat \Pi   ~=~ \cos \beta   \big(\cos \beta\,   \oneone  -  \sin \beta \, \sigma_1 \Gamma^{\hat z \psi}  \big)  \, \widehat  \Pi_{D1}  ~+~  \sin \beta   \big(\sin \beta\,   \oneone  +   \cos \beta \, \sigma_1 \Gamma^{\hat  z \psi}  \big)  \,  \widehat \Pi_{D5}  \,.  \label{BubbProj5a} 
\end{equation}
Note  that  $\beta$ and $\alpha$ are independent rotation angles. 

While well motivated, this form of the projector has not been rigorously established, 
because we have applied the equation for the new projector of a supertube transition (given in equation (\ref{genPistAnsatz}))
to two non-commuting projectors $\widehat \Pi_{D1}$ and $\widehat \Pi_{D5}$, 
whereas up to now the projectors were built out of commuting projectors.
In the Appendix we present a more detailed analysis of the system
in which we find its commuting projectors and obtain the \stratum\ projector rigorously; the final result is exactly 
the one in (\ref{BubbProj5a}).

One can now expand and simplify in a number of ways.  One such instructive form is:
\begin{equation}
\widehat \Pi   ~=~   {1 \over 2} \, \Big\{ \oneone + \Gamma^{0}  \big[\sin \alpha\,  \Gamma^{\hat \theta} ~+~ \cos\alpha\,  \Gamma^{\hat z}  \big(\cos \beta\,  \sigma_1 \oneone  +  \sin \beta \, \Gamma^{\hat z \psi}  \big) \big(\cos \beta\,   \oneone  -  \sin \beta \, \sigma_1 \Gamma^{1234 \hat z \psi}  \big)   \big]      \Big\} \,.  \label{BubbProj5b} 
\end{equation}
Again the second term in this equation squares to $\oneone$ and is
traceless, and hence $\widehat \Pi $ preserves sixteen supersymmetries.
In addition, (\ref{BubbProj5a}) shows that the nullspace of $\widehat
\Pi$ contains the common nullspace of $\widehat \Pi_{D1}$ and $\widehat
\Pi_{D5}$ while (\ref{BubbProj2c}) and (\ref{BubbProj3c}) show that this
common nullspace contains the common nullspace of $\Pi_{D1}$, $\Pi_{D5}$
and $\Pi_{P}$.  In other words, the sixteen supersymmetries preserved by
$\widehat \Pi $ contain the four supersymmetries of the D1--D5--P system
and these four supersymmetries are independent of the tilt angles, $(\alpha,
\beta)$, and of coordinates, $(\theta, \psi)$.

The \stratum\ is therefore a two-dimensional sheet swept out by $(\hat z,
\psi)$ and locally preserves the sixteen supersymmetries defined by
$\widehat \Pi$.  However as the directions $\theta$ and $\psi$ vary
there is always a set of four supersymmetries common to all the local
pieces of the \stratum\ and so all these pieces are mutually BPS and
non-interacting and so one should be able to assemble them into a
complete $\frac{1}{8}$-BPS \stratum\ that has an arbitrary
two-dimensional shape.

We have thus completed the local argument that strongly suggests that our
conjectured \stratum\ should exist as a regular solution in string
theory.

\section{Conclusions}
 \label{Conclusions}

We have shown that there should exist a completely new set of \nBPS{8} bound states of D-branes, namely, \strata.  These have three electric charges and globally preserve $4$ supersymmetries while locally appear to be \nBPS{2} objects preserving $16$ supersymmetries.  Moreover, they have a KKM dipole charge whose world-volume wraps a codimension-three surface, and hence their back-reaction should yield smooth supergravity solutions. The shape of the \strata\ in five dimensions is determined by five functions of \emph{two variables}, and since \strata\ have the same charges as the D1-D5-P black hole their back-reaction will describe microstates of this black hole in the same regime of parameters where the black hole exists.  The fact that these  microstate geometries depend on functions of two variables leads us to expect that they will be able to account for vastly more entropy than ordinary supertubes, whose shapes only depend upon functions of one variable.

The new bound states should describe particular degrees of freedom of the D1-D5-P system, which one may hope to be able to see either by studying the appropriate microscopics or by relating normalizable modes of the supergravity solution, when it is constructed, to the various expectation values of the dual theory (as was done for simpler systems in \cite{Alday:2005xj,Kanitscheider:2006zf}). While the second method may indeed yield interesting physical information, it is unlikely that one will be able to describe microscopically the \strata\ using the non-Abelian degrees of freedom of the D1-D5-P system, essentially because one of the dipole charges corresponds to a KK-monopole, and describing systems with dipole charges whose tension scales like $1/g_s^2$ (KK-monopoles of NS5 branes) using brane non-Abelian degrees of freedom is equivalent to proving confinement \cite{Polchinski:2000uf}.

While the arguments in this paper strongly suggest that  \strata\ exist, there are still some serious calculations to be done to prove their existence. Of course, ideally one should construct a fully-back-reacted supergravity solution that depends on five functions of two variables and has the charges and the dipole charges indicated in this paper. 
In fact, two special limits of this would-be solution have already been constructed in the literature. As we explained in Section \ref{physdesc} and illustrated in Fig.~\ref{superstratum}, the \strata\ can be thought as coming from a smooth  D1-D5 supertube with a KKM dipole charge to which one adds momentum-carrying  modes that  break the isometry along the common D1-D5 direction. A perturbative solution in which \emph{one} unit of momentum is added to the smooth D1-D5 supertube was constructed in \cite{Mathur:2003hj}, and the rather non-trivial matchings that insured the existence of that solution make us confident that more complicated \stratum\ solutions exist.

A second highly-non-trivial supergravity solution that can be thought of as a limit of a \stratum\ was obtained in \cite{Ford:2006yb} by spectrally-flowing a supertube of arbitrary shape. The resulting solution depends non-trivially both on the common D1-D5 direction\footnote{Consequently this represents  a supersymmetric solution of six-dimensional ungauged supergravity \cite{Gutowski:2003rg} that does not descend to five dimensions.} and on one of the angles in the base, but this dependence is ``the same'' in that this solution is a \stratum\ whose function of two variables only depends on their sum but not on their difference.  Given the existence of these non-trivial limit solutions, and given that the physical description provided in this paper gives a rather precise description of its charges and dipole moments, we believe the complete construction of the supergravity \stratum\ solution to be within reach. 

There are, however, some potential issues that might arise in this construction.  First, our local argument is based upon the fact that BPS bits of the stratum will not interact, since they are mutually BPS, and so can be assembled at will into the \stratum.  This is not exactly true:  multi-charge BPS configurations \emph{do} interact if they are not mutually local and must satisfy bubble equations or integrability conditions that constrain their locations. However, we do not expect this to be a problem because there is still freedom to adjust some  of the electric charge and angular momentum densities so as to accommodate the freedom to adjust the relative locations of the bits of \stratum.  Put differently, we expect any such conditions not to emerge as restrictions on the shape but to constrain, for a given shape, certain integrals of the charge and angular-momentum densities (much as one finds for wiggly supertubes in  bubbling solutions \cite{Bena:2010gg}).

Another delicate issue that might constrain the \strata\ is the fact that one must compactify the common D1-D5 direction, $\hat z$, and have a properly quantized coefficient of the potential along this $U(1)$ fiber in order to smooth out the geometry using the KK monopole.  This works beautifully for the usual D1-D5 configuration where $\hat z$ is simply the coordinate along a compactification circle.  For \strata, the coordinate $\hat z$ parametrizes a curve of arbitrary shape and so there may be an issue in making the KKM construction smooth on such a geometry.  Once again we suspect that this will not present a problem because  the process of adding charges and smearing them out includes some choices of charge density functions and this geometric issue should simply amount to selecting the KKM density distribution so that it is a fixed integer along the curve defined by $\hat z$.   

We raise these issues to show that the complete proof of the existence of \strata\ as microstate geometries  still requires some further work.   Indeed, the supergravity solutions corresponding to \strata\ will  be extremely interesting.  One should recall that, from the six-dimensional perspective, the smooth geometry created by the D1-D5 supertube is simply a non-trivial cycle in the  three-dimensional homology of the space-time and that the usual supertube profile represents fluctuations of this $3$-cycle that depend upon functions of one variable.  The \stratum\ will thus represent the much richer and more extensive space of two-dimensional fluctuations of this $3$-cycle. We are thus very  optimistic about these new solutions and the role that we expect them to  play in black-hole thermodynamics.  

The local construction of the \strata\ has revealed some particularly satisfying properties. First, they seem to be the most natural fundamental bound-state constituent of a three-charge black hole.  The two-charge supertube is a simple fusion of two electric charges to form a fundamental bound object and much of the work of Mathur and collaborators has shown how these objects naturally carry the entropy of the two-charge system.  Once one has a three-charge system we now find that there is a very natural ``double bubbling'' that leads to a fundamental, bound object that carries all three charges and by very good fortune these configurations depend upon functions of two variables.  While supertubes and generalized supertubes may account for some of the entropy of the three-charge black hole, they are not really fundamentally three-charge objects whereas the new \stratum\ is precisely such an object and should carry far more entropy\footnote{A generalization of the ``free supergravity estimate'' of  \cite{deBoer:2009un} to the D1-D5-P system suggests that the entropy that one can maximally obtain from supergravity should behave as $\sim c^{3/4} L_0^{1/4}$ for large $L_0$ (using two-dimensional CFT notation), which would still be smaller than the Cardy result $\sim c^{1/2} L_0^{1/2}$. It will be interesting to see whether \strata\ respect this entropy bound, or can evade it because of their non-perturbative dipole charges.}.  We find it remarkable that such an object emerges precisely when the three-charge entropy seems to require it and its doubly remarkable that this object can be represented in terms of a microstate geometry.

In this paper, we studied the possibility of a particular two-fold
supertube transition of the D1-D5-P system.  Namely, at the first stage
of the transition, the D1 and P charges are transformed into a tilted D1-P system and the D5
and P charges into a tilted D5-P system. Then, at the second stage, these tilted
D1-P and D5-P systems are puffed up into a \stratum.  However, since the D1, D5 and P charges
can be dualized into each other,  at the first stage we
could have also considered the possibility of the D1 and D5 charges 
forming a supertube with KKM dipole charge, which in turn could have also participated in the
second transition. It is not clear whether such different patterns of
two-fold supertube transitions give the same final result, and if not, whether the 
resulting configuration could still source a smooth geometry. 
We leave further investigations into such dynamical issues for future research.



\bigskip
\leftline{\bf Acknowledgements}
\smallskip

\noindent We would like to thank S. Giusto, K. Papadodimas, S. Raju and
C. Ruef for helpful discussions. The work of IB was supported in part by
the ANR grant 08-JCJC-0001-0, and by the ERC Starting Independent
Researcher Grant 240210 - String-QCD-BH\@. The work of NPW was supported
in part by DOE grant DE-FG03-84ER-40168.  This work was partly supported
by the Foundation of Fundamental Research on Matter (FOM). MS is very
grateful to the ITFA, University of Amsterdam for hospitality.  MS and
NPW are also very grateful to the IPhT, CEA-Saclay for hospitality while
much of this work was done.



\appendix
\bigskip
\bigskip
\bigskip
\leftline{\bf\Large Appendix}
\section{Detailed analysis of the projectors for double bubbling}
\label{appendixA}

In this appendix, we develop and study the projectors for double bubbling in  detail. Most of the arguments about the supersymmetry of supertubes in various duality frames can be found in the original papers \cite{Mateos:2001qs,Mateos:2002yf, Lunin:2001fv}, but here we try to give a general picture that is independent of the duality frame used to describe the supertube. 
In \ref{apsc:susyalg}, we summarize the \nBPS{2} projectors for various
branes and solitons and explain, via examples, how to combine them to
construct BPS states with given charges.  In
\ref{apsc:genericSupertube}, we derive the projector \eqref{genPist} for
generic supertube transitions where a combination of two electric
charges transforms into a new configuration with new dipole charge.  In
\ref{apsc:doubleBubbling}, we use this result to construct the projector
for the double bubbling, namely the two-fold supertube transition.  In
\ref{apsc:theRelation}, we explain the relation between the projector
derived here and the one used in the main text.

Our convention for the ten-dimensional Clifford algebra are that $\{\Gamma^\mu,\Gamma^\nu\}=2\eta^{\mu\nu}$, wih
$\eta^{\mu\nu}=(-+\dots +)$ and we define $\Gamma^{\mu_1\dots \mu_k}\equiv\Gamma^{[\mu_1}\cdots \Gamma^{\mu_k]}$.  We also define  the ten-dimensional helicity operator by $\Gamma_*\equiv\Gamma^{0\dots 9}$, with $\Gamma_*^2 = \oneone$.

In type II superstring theory, there are two ten-dimensional
Majorana-Weyl supercharges, $Q$ and $\Qt$, each of which will be
described in terms of $32$ component spinors satisfying Majorana and
helicity constraints.  In type IIA they have opposite helicity,
$\Gamma_* Q=Q$ and $\Gamma_* \Qt=-\Qt$, while in type IIB they have the
same helicity, $\Gamma_* Q=Q$ and $\Gamma_* \Qt=\Qt$.  We will also
think of these supersymmetries as belonging to a doublet:
$\cQ=\bigl(\begin{smallmatrix}Q\\ \Qt\end{smallmatrix}\bigr)$ that has
$2 \times 32 =64 $ components and the Pauli $\sigma$-matrices will be
thought of as acting on the doublet label $\cQ$. This means that the
helicity projector, $\Gamma_*$, is equivalent to the action of
$\sigma_3$ in the type IIA theory and  $\oneone_2$ in type IIB\@.  We
will understand that the $\sigma$ matrices and the identity matrix,
$\oneone_2$, are implicitly tensored with the action of the gamma
matrices, $\Gamma^\mu$, and that $\oneone_{32}$ is implicitly tensored
with the action of the $\sigma$-matrices.

\subsection{Supersymmetry algebra and projectors for branes}
\label{apsc:susyalg}

The supercharges preserved by various fundamental branes and solitons satisfy 
\begin{align}
 \Pi \cQ&=0, \qquad \Pi={1\over 2}(1+P),
\end{align}
where the matrices, $P$,  are given by \cite{Bena:2002kq, Grana:2002tu, Hassan:1999bv}:
\begin{align}
 P_{\rm P  }&=\Gamma^{01}&
 P_{\rm F1 }&=\Gamma^{01}\sigma_3\notag\\
 P_{\rm NS5}^{\rm IIA}&=\Gamma^{012345}&
 P_{\rm NS5}^{\rm IIB}&=\Gamma^{012345}\sigma_3\notag\\
 P_{\rm KKM(12345;6)}^{\rm IIA}&=\Gamma^{012345}\sigma_3 = \Gamma^{6789}&
 P_{\rm KKM(12345;6)}^{\rm IIB}&=\Gamma^{012345} =  \Gamma^{6789}\notag\\
 P_{\rm D0 }&=\Gamma^0 i\sigma_2&
 P_{\rm D1 }&=\Gamma^{01} \sigma_1\notag\\
 P_{\rm D2 }&=\Gamma^{012} \sigma_1&
 P_{\rm D3 }&=\Gamma^{0123} i\sigma_2\notag\\
 P_{\rm D4 }&=\Gamma^{01234} i\sigma_2&
 P_{\rm D5 }&=\Gamma^{012345} \sigma_1\notag\\
 P_{\rm D6 }&=\Gamma^{0123456} \sigma_1
 \label{variousProjectors}
\end{align}

From these conditions, one can reverse-engineer the supersymmetry algebra.  For example, one can show that the F1 condition in \eqref{variousProjectors} means that we have the following terms on the right hand side of the $\cQ,\cQ^\dagger$ anticommutator: 
\begin{align}
 {1\over 2}\{\cQ,\cQ^\dagger\}=
 P_\mu\Gamma^{0\mu}+\tau_{F1}Q^{F1}_\mu\Gamma^{0\mu}\sigma_3\,,
 \qquad\qquad
\tau_{F1}={1\over 2\pi \alpha'},
\label{gci3Aug10}
\end{align}
where $Q_\mu^{F1}$ corresponds to the charge of fundamental strings.

One can show this as follows.  If we have straight fundamental strings at rest, the supersymmetry
algebra \eqref{gci3Aug10} becomes
\begin{align}
 {1\over 2}\{\cQ,\cQ^\dagger\}= M+q_i\Gamma^{0i}\sigma_3,\qquad \tau_{F1}Q^{F1}_\mu\equiv (0,{\bf q})\,
\label{susyalg1}
\end{align}
The charge vector ${\bf q}$ measures the tension of the fundamental string, including the direction and multiplicity.  Now, assume that the supercharge $\cQ$ also satisfies:
\begin{align}
 \Pi\cQ=0,\qquad \Pi
 ={1\over 2}\left(1+{q_i\over |{\bf q}|}\Gamma^{0i}\sigma_3\right)\,, 
\label{hpe3Aug10}
\end{align}
corresponding to fundamental strings along the direction ${\bf q}\over |{\bf q}|$.  Equivalently, one has:
\begin{align}
 \Pi'\cQ=\cQ,\qquad \Pi'   \equiv 1-\Pi
 ={1\over 2}\left(1-{q_i\over |{\bf q}|}\Gamma^{0i}\sigma_3\right).\label{hox3Aug10}
\end{align}
The superchage $\cQ$ satisfies the following relation:
\begin{align}
 {1\over 2}\{\cQ,\cQ^\dagger\}
 &= {1\over 2}\{\Pi'\cQ,(\Pi'\cQ)^\dagger\}
 = {1\over 2}\Pi'\{\cQ,\cQ^\dagger\}\Pi'\notag\\
 &= {1\over 4}
\left(1-{q_i\over |{\bf q}|}\Gamma^{0i}\sigma_3\right)
 (M+q_i\Gamma^{0i}\sigma_3)
\left(1-{q_i\over |{\bf q}|}\Gamma^{0i}\sigma_3\right)
 =(M-|{\bf q}|)\Pi',\label{heb27May11}
\end{align}
which vanishes for a BPS configuration of mass $M=|{\bf q}|$. So, the supercharge, $\cQ$, satisfying \eqref{hox3Aug10} is preserved in this F1 configuration.
As one can see from (\ref{hpe3Aug10}), half the eigenvalues of $\Pi$ are $1$ and the other half are $0$ and
so half  the components of $\cQ$ survive the projection \eqref{hpe3Aug10}, and hence the state satisfying \eqref{gci3Aug10} is \nBPS{2}.

The essential point is that (\ref{susyalg1}) becomes the projector (\ref{hpe3Aug10}) precisely on the BPS states and, conversely the supersymmetry algebra must be compatible with the projectors that define \nBPS{2} states. 

In this manner, one can determine  the supersymmetry algebra for more general configurations to be:
\begin{align}
 {1\over 2}\{\cQ,\cQ^\dagger\}&=
 \begin{cases}
 P_\mu\Gamma^{0\mu}+\tau_{F1}Q^{F1}_\mu\Gamma^{0\mu}\, \sigma_3
 +\tau_{D1}Q^{D0} \Gamma^{0} \, i\sigma_2
 +\tau_{D2}Q^{D2}_{\mu_1\mu_2} \Gamma^{0\mu_1\mu_2} \, \sigma_1
 +\cdots&\text{(IIA)}\label{susyalgIIAB}
 \\
 P_\mu\Gamma^{0\mu}+\tau_{F1}Q^{F1}_\mu\Gamma^{0\mu}\, \sigma_3
 +\tau_{D1}Q^{D1}_\mu \Gamma^{0\mu} \, \sigma_1
 +\tau_{D3}Q^{D3}_{\mu_1\mu_2\mu_3} \Gamma^{0\mu_1\mu_2\mu_3}\,  i\sigma_2
 +\cdots&\text{(IIB)}
\end{cases}
\end{align}
where $\tau_{Dp}=(2\pi)^{-p}(\alpha')^{-(p+1)/2} g_s^{-1}$.

Another instructive exercise is to derive the linear combination of supersymmetries that vanishes on states with particular combinations of charges. This then describes  the BPS bound states with those  charges.  For a bound state of F1(1) and D1(1) at rest\footnote{As in the main body of this paper,  the parentheses after a configuration label denotes the spatial directions they wrap, for example,  F1($i$) means a  fundamental string wrapped along $x^i$.}, the anticommutator, \eqref{susyalgIIAB}, is simply:
\begin{align}
 {1\over 2}\{\cQ,\cQ^\dagger\}=
 M+\tau_{F1}Q^{F1}_1\Gamma^{01}\sigma_3
  +\tau_{D1}Q^{D1}_1\Gamma^{01}\sigma_1.
 \label{hhy27May11}
\end{align}
Just as we saw in \eqref{heb27May11}, it is easy to show that \eqref{hhy27May11} vanishes if $\cQ$ satisfies
\begin{align}
 \Pi_{\rm F1D1}\cQ=0,\qquad
 \Pi_{\rm F1D1}={1\over 2}\left[1+\Gamma^{01} (\cos\beta\,\sigma_3+\sin\beta\, \sigma_1)\right],
 \qquad
 \tan\beta={\tau_{D1}Q^{D1}_1\over \tau_{F1}Q^{F1}_1},
 \label{jvcm22Jul10}
\end{align}
provided that the mass is equal to
$M=\sqrt{(\tau_{F1}Q^{F1}_1)^2+(\tau_{D1}Q^{D1}_1)^2}$.  The angle
$\beta$ is a ``mixing angle'' between F1 and D1.  It is again easy to
see that this F1-D1 bound state is \nBPS{2}.

Similarly, a D1 brane with charge vector ${\bf q}$, boosted transverse to its world-volume
with momentum $\bf p$, such that  ${\bf q}\cdot{\bf p}=0$,  is also \nBPS{2}, and satisfies
\begin{align}
 \left(1+{p_i\over M}\Gamma^{0i}+{q_i\over M}\Gamma^{0i} \sigma_1\right)\cQ=0,
 \qquad  M=\sqrt{{\bf p}^2+{\bf q}^2}.\label{jxlt4Aug10}
\end{align}
The mass, $M$, obtained from this BPS condition is indeed the mass of an object with rest mass $|\bf q|$ boosted to have momentum $\bf p$.

We can also derive this result by directly boosting a D1-brane in a
transverse direction. Starting with the projector $\Pi_{\rm
D1}=\frac{1}{2}(1+\Gamma^{01}\sigma_1)$ for a D1-brane in the $x^1$
direction, boosting in the $x^2$-direction amounts to replacing
\begin{equation}
 \label{auxsub}
\Gamma^0 \rightarrow \cosh \xi \Gamma^0 - \sinh\xi \Gamma^2.
\end{equation}
The resulting projector does not look like (\ref{jxlt4Aug10}) but can be
brought in that form as follows. First multiply $\Pi_{\rm D1}$ on the
left by $\Gamma^0$, then perform the substitution (\ref{auxsub}), and
finally multiply the projector once more on the left by
$-\Gamma^0/\cosh\xi$. The result is precisely of the form
(\ref{jxlt4Aug10}) with $p/M=\tanh\xi$ and $q/M=1/\cosh\xi$.

The examples above were all \nBPS{2} states.  Configurations with less supersymmetry can be studied in precisely the same manner.  For example, if we have D1(1) and D5(16789), the supersymmetry algebra has:
\begin{align}
 {1\over 2}\{\cQ,\cQ^\dagger\}=
 M+q_1\Gamma^{01}\sigma_1+q_5\Gamma^{016789}\sigma_1.\label{muov2Aug10}
\end{align}
The preserved supersymmetry, $\cQ$, satisfies
\begin{align}
 \Pi_1\cQ=\Pi_5\cQ=0\label{itvq2Aug10}
\end{align}
where
\begin{align}
 \Pi_1={1\over 2}(1+\Gamma^{01}\sigma_1),\qquad
 \Pi_5={1\over 2}(1+\Gamma^{016789}\sigma_1),\qquad
 [\Pi_1,\Pi_5]=0\label{projD1D5}
\end{align}
and we assumed that $q_1,q_5>0$.  Because $\Pi_1$ and $\Pi_5$ commute, the two conditions in \eqref{itvq2Aug10} are equivalent to the single condition
\begin{align}
 \Pi_{15} \cQ=0,\qquad \Pi_{15}=1-(1-\Pi_1)(1-\Pi_5)
 ={1\over 4}(3+\Gamma^{01}\sigma_1-\Gamma^{6789}+\Gamma^{016789}\sigma_1),
\end{align}
or, equivalently,
\begin{align}
 \Pi_{15}' \cQ=\cQ,\qquad \Pi_{15}'=1-\Pi_{15}=\Pi_1'\Pi_5'
 ={1\over 4}(1-\Gamma^{01}\sigma_1+\Gamma^{6789}-\Gamma^{016789}\sigma_1).\label{pqd3Aug10}
\end{align}
Now, from the supersymmetry anticommutator, \eqref{muov2Aug10}, one can derive
\begin{align}
 {1\over 2}\{\cQ,\cQ^\dagger\} = {1\over 2}\Pi'_{15}\{\cQ,\cQ^\dagger\}\Pi'_{15}
&=  (M-q_1-q_5)\Pi'_{15}\,. 
 \end{align}
This vanishes for the BPS mass, $M=q_1+q_5$.  The fact that $\tr(\Pi_{15}')={1\over 4}\tr(\oneone)$ means that this is a
\nBPS{4} state.

\subsection{The generic supertube transition}
\label{apsc:genericSupertube}

A general supertube transition takes the form:
\begin{equation}
\left( \begin{array}{c}  Q_1 \, (x) \\ Q_2\,  (y)  \end{array}  \right) ~\rightarrow~ \left( \begin{array}{c}  d_1\,   (z\psi)\\  \cJ\,  (\psi) \end{array}  \right)
  \,, \label{GenSTnotnApp}
\end{equation}
where $x,y,z$ and $\psi$ are some subsets of coordinate directions along which the branes are wrapped or the momentum is directed.  The coordinate $\psi$ indicates the new direction associated with the supertube.
The supersymmetry preserved by the configuration after the transition depends upon the orientation of  the $\psi$ direction as one goes along the brane.  If we zoom in on a point on the supertube, the tube can be locally thought of as straight and we can determine the supersymmetry preserved at that point by the methods above.  The supertube has the following properties: (i) $16$ supersymmetries are preserved at each point along $\psi$, and (ii)  these  $16$  symmetries differ from point to point but they all share a common subset of eight supersymmetries and these are the same eight supersymmetries that are preserved by the original charge configuration before the transition.   From these requirements, it is straightforward to derive a general formula for the supersymmetry projector after the supertube transition for general charges $Q_1,Q_2$ in \eqref{GenSTnotnApp}.

\subsubsection{An example: the F1-P system}
\label{sssapp:F1P}

In order to derive the supersymmtry projector for the general supertube transition \eqref{GenSTnotnApp}, we begin with a simple example, the F1-P system, where we know the detailed physics of the supertube transition \cite{Lunin:2001fv} and can construct the projector from the knowledge of the configuration.   We will also take a different approach to the one in the main body of this paper.

Consider the F1-P system in which $w$ fundamental strings are wound along $z$ and $n$ units of momentum are directed along the same $z$
direction. We take $z$ to be a compact direction with periodicity $2\pi
R_z$.  The projectors associated with these charges are:
\begin{align}
 \Pi_{F1(z)}={1\over 2}(1+\Gamma^{0z}\sigma_3),\qquad
 \Pi_{P(z)}={1\over 2}(1+\Gamma^{0z}),\qquad
 [\Pi_{F1(z)},\Pi_{P(z)}]=0.
 \label{proj-F1(z)P(z)}
\end{align}
It is easy to see that supercharge $\cQ$ annihilated by $\Pi_{F1(z)}$ and $\Pi_{P(z)}$ describes a \nBPS{4} state.

We now make wish to make a supertube transition:
\begin{align}
 \label{hxz5Aug10}
 \begin{matrix}w\\ n \end{matrix}
 \begin{pmatrix}F1(z)\\ P(z) \end{pmatrix}
 ~~\to ~~
 \begin{matrix}d\\ J \end{matrix}
 \begin{pmatrix}f1(\psi)\\ \cJ(\psi) \end{pmatrix}.
\end{align}
%
\begin{figure}
\begin{center}
 \includegraphics[height=4.5cm]{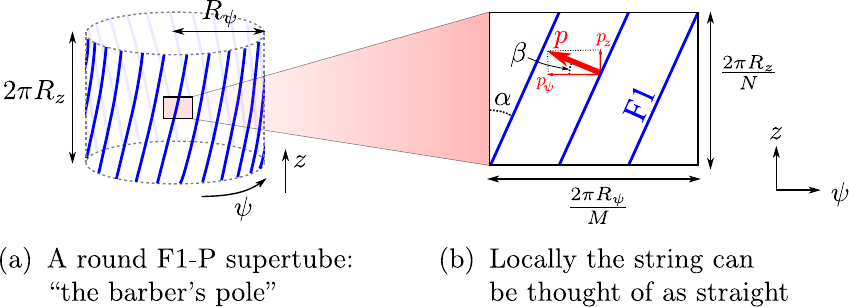}
 \caption{The zoom-in near an F1-P supertube profile}\label{f1p-tube}
\end{center}
\end{figure}%
%

For this system, the fundamental bound state is a string with a momentum wave on it \cite{Dabholkar:1995nc}, which can be thought of as the result of a supertube transition that adds  an extra F1 dipole charge and an extra angular momentum along the $\psi$ direction \cite{Lunin:2001fv}.  In the new bound state the string world-sheet extends along a curve parametrized by $\psi$ as a function of the original world-sheet direction, $z$, and carries $J$ units of momentum along $\psi$. In particular, if $\psi = z \tan \alpha $ 
the shape of the string is a helix moving up (or down, depending the signs of charges) along its axis, just like the barber's pole, as depicted in Fig.~\ref{f1p-tube}a.  By studying such circular
traveling waves on strings using the Nambu--Goto action or by looking at
the corresponding supergravity solutions \cite{Lunin:2001fv}, one finds
that the charges $w,n$ and
the dipole charges $d,J$ satisfy the relation:
\begin{align}
\label{nw=Jd_F1}
 nw=Jd \,,
\end{align}  
and that the angular momentum $J$ and the radius of the $\psi$ circle are related by:
\begin{align}
 J=2\pi R_\psi^2 \tau_{F1} d\,. \label{Rpsi} 
\end{align}  

Note that this relation implies that the local ``angular momentum density'' (proportional to $J/R_\psi^2$)\footnote{Technically the angular momentum density is $J/R$ while the quantity $J/R^2$ is the linear momentum density around the supertube, or simply the angular speed.  Thus what we refer to as ``angular momentum density,'' $\rho_J$, is really the momentum density {\it along} the tube and should perhaps be more consistently denoted by $\rho_p$, but this could lead to notational confusion with $\rho_P$ and so we will persist with the mild abuse of terminology in calling this quantity an angular momentum density and denoting it by $\rho_J$.}, must be equal to the local dipole charge density, and hence be constant along the profile; we discuss this in more detail below (around equation \eqref{irb10Jul11}).
This relation can also be understood as coming from requiring the Killing spinors of the supertube bit to be the same as those of its electric (F1 and P) charges. Adding dipole charges shifts these Killing spinors, and so does adding angular momentum; however if \eqref{Rpsi} is satisfied the shifts cancel each other and the Killing spinors become again those of the electric charges. This can also be seen from supergravity analysis of the near-supertube solution \cite{Bena:2004wv}.

Consider a very small part of the helix, as shown in Fig.~\ref{f1p-tube}b. Let the size of the square we are focusing on be $(2\pi R_\psi/ M)\times (2\pi R_z / N)$ for some very large integers $M$ and $N$.  We can think of the strings in this small square as straight lines in a small region on the $(\psi,z)$-plane.
Let the angle between this string and the $z$ axis be $\alpha$ (see Fig.\ \ref{f1p-tube}b).  Because
the string is wrapped $d/N$ times along the $\psi$ direction of length
$2 \pi R_\psi/M$ and $w/M$ times along the $z$ direction of length $2
\pi R_\psi/N$, the angle $\alpha$ satisfies
\begin{align}
\label{tanalpha}
 \tan\alpha
={{d\over N}\cdot {2\pi R_\psi\over M}\over {w\over M}\cdot {2\pi R_z\over N}}
 ={d \,R_\psi \over w R_z }
 =\sqrt{n/R_z\over 2\pi w R_z \tau_{F1}}
,
\end{align}
where in the third equality we used the relations \eqref{nw=Jd_F1} and
\eqref{Rpsi}.  


Let the angle between the momentum vector carried by the string and the
$\psi$ axis be $\beta$ (see Fig.\ \ref{f1p-tube}b). The $(\psi,z)$
components of the momentum carried by the entire helix is
$(-J/R_\psi,n/R_z)$.  So, the momentum carried by the piece of strings
in the small square is $(p_\psi,p_z)=(-{J\over MN
R_\psi},{n\over MN R_z})$.  So, the angle, $\beta$, between the momentum vector carried
by the string bit and the $\psi$ axis satisfies:
\begin{align}
 \tan\beta
 ={{n/(MNR_z)}\over J/(MNR_\psi)}
 ={nR_\psi\over JR_z}
 =\sqrt{n/R_z\over 2\pi w R_z \tau_{F1}}.
 \label{tanbeta}
\end{align}
In the last equality we used the relations required by the dynamics of
the string, \eqref{nw=Jd_F1} and \eqref{Rpsi}. Comparing this with
\eqref{tanalpha} we see that $\beta=\alpha$ and hence these dynamical
conditions require that the momentum be perpendicular to the direction
of the string, {\it i.e.}, the string is boosted transverse to its
world-sheet.

\begin{figure}
\begin{center}
 \includegraphics[height=4cm]{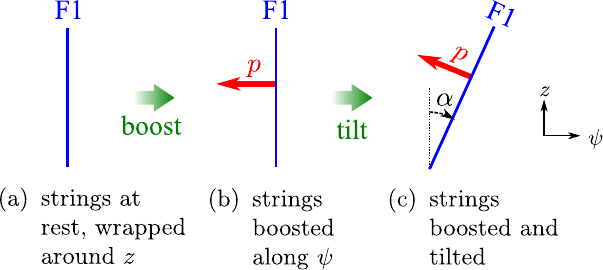} \caption{Boosting and
 rotating to obtain the desired F1-P
 configuration}\label{boost-n-rotate}
\end{center}
\end{figure}
%
Based on this physical picture, we will construct the supersymmetry projector for this configuration.  Starting from strings sitting at rest and extending along $z$ (Fig.~\ref{boost-n-rotate}a), we boost them along the negative $\psi$-direction (Fig.~\ref{boost-n-rotate}b). We then rotate them in the $(\psi,z)$ plane by angle $-\alpha$ to get to the desired configuration (Fig~\ref{boost-n-rotate}c).
The projector for F1($z$) boosted along the negative $\psi$ direction (Fig.~\ref{boost-n-rotate}b) is given by:
\begin{align}
 \Pi_{\text{F1(z)-P($\psi$)}}
 =  {1\over 2}\left(1-\sin\gamma\, \Gamma^{0\psi}+\cos\gamma\,\Gamma^{01} \sigma_3\right).
 \label{projF1(z)P(psi)}
\end{align}
Here, $\gamma$ is related to the ratio of the P($\psi$) charge and the
F1($z$) charge as
\begin{align}
 \tan\gamma
 &={\text{(momentum P($\psi$))}\over \text{(F1($z$) charge)}}
 \quad\text{for the configuration of Fig.\ \ref{boost-n-rotate}b}
.
\end{align}
The magnitudes of F1 and momentum charges in Fig.\ \ref{boost-n-rotate}b
are equal to those in the configuration shown in Fig.\
\ref{boost-n-rotate}c, because they are related to each other by
rotation. Therefore,
\begin{align}
 \tan\gamma
 &={\sqrt{(p_\psi)^2+(p_z)^2}\over \tau_{F1}\cdot\text{(length of F1)}}
 ={\sqrt{\bigl({1\over MN}{n\over R_z}\bigr)^2+\bigl({1\over MN}{J\over R_\psi}\bigr)^2}
 \over
 \tau_{F1}\sqrt{\bigl({w\over M}{2\pi R_z \over N}\bigr)^2
 +\bigl({d\over N}{2\pi R_\psi \over M}\bigr)^2}}
 ={\sqrt{({n/R_z})^2+({J/R_\psi})^2}
 \over
2\pi  \tau_{F1}\sqrt{({wR_z})^2
 +({dR_\psi})^2}}\,.
 \label{tangam1}
\end{align}
Using the relations \eqref{nw=Jd_F1} and \eqref{Rpsi}, it is easy to
show that this is equal to \eqref{tanalpha} and thus
$\gamma=\alpha=\beta$.

By rotating the projector \eqref{projF1(z)P(psi)} (with $\gamma=\alpha$) by angle $-\alpha$ in the $(\psi,z)$-plane, we get the
projector after the supertube transition:
\begin{align}
  \Pi_{st}
 &=
 {1\over 2}
 \Bigl[1-s(c\Gamma^{0\psi}-s\Gamma^{01})
 +c(c\Gamma^{01}+s\Gamma^{0\psi}) \sigma_3\Bigr],
\label{proj-F1P-st-1}
\end{align}
where $s=\sin\alpha$ and $c=\cos\alpha$. The fact that
$\tr(\Pi_{st})={1\over 2}\tr(\oneone)$ means that this projector
preserves 16 supersymmetries.  These supersymmetries depend upon the
angle $\alpha$ as well as  the $\psi$ direction in space and hence
the position along the string.  However, note that \eqref{proj-F1P-st-1}
can also be written as:
\begin{equation}
 \Pi_{st}  = c(c+s\Gamma^{z\psi}\sigma_3)  \Pi_{F1(z)}   +  s(s-c\Gamma^{z\psi}\sigma_3  )  \Pi_{P(z)}.
 \label{proj-F1P-st2b}
\end{equation}
This implies that the projector $\Pi_{st}$ always preserves the set of
eight supersymmetries that are preserved by the original two projectors
$\Pi_{F1(z)}$ and $\Pi_{P(z)}$, independent of position along the $\psi$
curve.

Although we started by working with a circular $\psi$ curve, our local
analysis is valid also for an arbitrary curve parametrized by $\psi$
embedded in the $\mathbb{R}^8$ transverse to the original direction, $z$.
Such a curve in $\mathbb{R}^8$ can be parametrized by seven functions.
Furthermore, we can change the angle $\alpha$ along the curve.  Note
that $\alpha$ can be expressed in terms of the local charge densities, as
follows.  The momentum along $z$ carried by the part of the string in
our small square is
\begin{align}
 p_z=
{n\over R_z}\cdot{ 1\over MN}
={n\over M N R_z}.
\end{align}
So, the local momentum density per unit area is
\begin{align}
 \rho_{P}^{}= {p_z\over (2\pi R_z/N)(2\pi R_\psi/M)}
 ={n\over (2\pi R_z)^2R_\psi}
\end{align}
Similarly, the F1($z$) charge carried by the same string bits and the
local F1($z$) charge density are computed as
\begin{align}
 q_{F1}^{}
 =\tau_{F1}^{}\cdot 2\pi R_z w\cdot {1\over M N}
 ={2\pi w \tau_{F1}^{} R_z\over M N},\quad
 \rho_{F1}^{}= {q_{F1}^{}\over (2\pi R_z/N)(2\pi R_\psi/M)}
 ={w\tau_{F1}^{}\over 2\pi R_\psi}.\label{kagi7Jun11}
\end{align}
Comparing these with \eqref{tanalpha}, it is easy to see that the angle
$\alpha$ can be written as
\begin{align}
\tan\alpha=\sqrt{\rho_{P}^{}\over \rho_{F1}^{}}.\label{fezk7Jun11}
\end{align}
So, varying $\alpha$ corresponds to varying the ratio of charge
densities along the curve.  However, one cannot change all the charge densities completely freely:  It is easy to show that the dynamical
relation \eqref{nw=Jd_F1} implies the following relation:
\begin{align}
 \rho^{}_{P}\rho_{F1}^{}
 =\rho_{\cJ}^{}\rho_{f1}^{},\label{itz10Jul11}
\end{align}
where
\begin{align}
 \rho^{}_{\cJ}={J\over (2\pi R_\psi)^2R_z}={d\,\tau_{F1}^{}\over 2\pi R_z},\qquad
 \rho^{}_{f1}={d\,\tau_{F1}^{}\over 2\pi R_z}\label{irb10Jul11}
\end{align}
are the P($\psi$) and F1($\psi$) densities.  

Because the dipole and the angular momentum densities $\rho^{}_{f1}$ and $\rho^{}_{\cJ}$ are constant along the
curve, the relation \eqref{itz10Jul11} means that the product of the
charge densities, $\rho^{}_{P}\rho^{}_{F1}$, should be constant along
the curve. Taking this constraint into account, we have a supertube
parametrized by $7+1=8$ functions in total.

\subsubsection{The formula for the projector}
\label{Projector-derivation}

Based on the previous example with projectors  \eqref{proj-F1P-st-1} and \eqref{proj-F1P-st2b},  we can construct the projector for the general supertube transition \eqref{GenSTnotnApp}.  Let the \emph{commuting} projectors for the original electric charges  corresponding to the left hand side of \eqref{GenSTnotnApp}) be:
\begin{align}
 \Pi_1={1\over 2}(1+P_1),\qquad
 \Pi_2={1\over 2}(1+P_2),\qquad
 [\Pi_1,\Pi_2]=0.\label{fyiv20Sep10}
\end{align}
Because $\Pi_{1,2}$ are projectors and they  commute, $P_{1,2}$ satisfy
\begin{align}
 P_1^2=P_2^2=1,\qquad\qquad [P_1,P_2]=0.\label{fyft20Sep10}
\end{align}
For each of $\Pi_{1,2}$ to preserve sixteen supersymmetries, and for them together to preserve eight supersymmetries, we require $\tr P_1=\tr P_2=\tr (P_1P_2)=0$.   Based on the expressions,  \eqref{proj-F1P-st-1} and \eqref{proj-F1P-st2b}, take the
following ansatz for the projector after the supertube transition:
\begin{align}
\begin{split}
  \Pi_{st}
 &={1\over 2}\Bigl[ 1+(c_1\Gamma^{0\psi}+s_1P_1)\Gamma^{0\psi}(c_2\Gamma^{0\psi}+s_2P_2) \Bigr],\\
&={1\over 2}\left(
 1+s_1c_2P_1+c_1s_2P_2+c_1c_2\Gamma^{0\psi}+s_1s_2P_1\Gamma^{0\psi}P_2
 \right)
\end{split}
\label{genPistAnsatz}
\end{align}
where $c_i=\cos\theta_i,$ $s_i=\cos\theta_i$, and $\theta_{i=1,2}$ are some angles to be determined. From the second expression in \eqref{genPistAnsatz}, we can see that the state annihilated by $\Pi_{st}$ has the original electric charges and momentum along $\psi$, which is the correct feature for the projector after the supertube transition.

The matrix $\Pi_{st}$ given in \eqref{genPistAnsatz} is not, in general, a projector.  However, if one has
\begin{align}
 \{\Gamma^{0\psi},P_1\}= \{\Gamma^{0\psi},P_2\}=0\label{GammaPAnticommute}
\end{align}
then it can be shown that $\Pi_{st}^2=\Pi_{st}$ and thus $\Pi_{st}$ is a projector.  We will assume \eqref{GammaPAnticommute} henceforth.  For  $\Pi_{st}$ to preserve $16$ supersymmetries, it is sufficient to assume that $\tr (\Gamma^{0\psi}P_1P_2)=0$.  In the example \eqref{proj-F1(z)P(z)}, where $P_1=\Gamma^{01}\sigma_3$ and $P_2=\Gamma^{01}$, all these conditions are indeed satisfied.

The most important requirement that our candidate projector \eqref{genPistAnsatz} should satisfy is that it annihilates the
supercharges annihilated by the original projectors \eqref{fyiv20Sep10}.  For this, note that \eqref{genPistAnsatz} can be brought to the following form:
\begin{align}
 \Pi_{st}={1\over 2}\Bigl[1-\sin(\theta_1+\theta_2)+\cos(\theta_1+\theta_2)\Gamma^{0\psi}
 +(\dots)\Pi_{1} +(\dots)\Pi_{2}\Bigr].
 \label{PistITOPi1Pi2}
\end{align}
To derive \eqref{PistITOPi1Pi2}, we first (anti)commute $P_i$ appearing in \eqref{genPistAnsatz} to the right of $\Gamma^{0\psi}$ and then re-express them in terms of $\Pi_i$.  So, for $\Pi_{st}$ to annihilate supercharges annihilated by $\Pi_i$, we need
\begin{align}
 \theta_1+\theta_2={\pi\over 2} \mod 2\pi.
\end{align}
If we set $\theta_1={\pi\over 2}+\alpha,\theta_2=-\alpha$, then
\eqref{genPistAnsatz} reduces to
\begin{align}
 \Pi_{st}
 &={1\over 2}\Bigl[ 1+(-s\Gamma^{0\psi}+cP_1)\Gamma^{0\psi}(c\Gamma^{0\psi}-sP_2) \Bigr]\notag\\
 &={1\over 2}\Bigl[ 1+c^2P_1+s^2P_2-sc\Gamma^{0\psi}+sc\Gamma^{0\psi}P_1P_2 \Bigr],
\label{genPist}
\end{align}
where $c=\cos\alpha,s=\sin\alpha$.

From the general formula \eqref{genPist}, we can read off how the system of two electric charges associated with $P_1$ and $P_2$ undergoes a transition into a supertube configuration with momentum along $\psi$ and a dipole charge, $d$, associated with $P_{dip}=\Gamma^{0\psi}P_1P_2$.  Note that, for a general supertube transition, this expression for $P_{dip}$ means that, unlike the F1-P system, the dipole component of the supertube configuration is not generically obtained from a tilt and boost of the original projectors.  
 
The supertube with F1-P charges  \eqref{proj-F1P-st-1} corresponds to taking
$P_1=\Gamma^{01}\sigma_3$, $P_2=\Gamma^{01}$,
$P_{dip}=\Gamma^{0\psi}\sigma_3$.  The supertube with D1-D5 charges in \eqref{Bubb1}
corresponds to taking $P_1=\Gamma^{0z}\sigma_1$,
$P_2=\Gamma^{01234z}\sigma_1$, $P_{dip}=\Gamma^{01234\psi}$.  The D1-P
and D5-P supertubes in \eqref{Bubb2} correspond to taking
$P_1=\Gamma^{0z}\sigma_1$, $P_2=\Gamma^{0z}$,
$P_{dip}=\Gamma^{0\psi}\sigma_1$ and $P_1=\Gamma^{01234z}\sigma_1$,
$P_2=\Gamma^{0z}$, $P_{dip}=\Gamma^{01234\psi}\sigma_1$, respectively.

Just as in the F1-P example, the supertube can be along an arbitrary curve transverse to the direction of the original electric charges.  Along such an arbitrary curve, one can also vary the angle, $\alpha$,  which is related to the ratio of the densities of the original electric charges via:
\begin{align}
 \tan\alpha&=\sqrt{\rho^{}_2\over \rho^{}_1} \,, \label{jecc2Jun11}
\end{align}
where $\rho^{}_{1}$ and $\rho^{}_{2}$ are the densities of the original electric charges. Equation \eqref{jecc2Jun11} is a generalization of the relation \eqref{fezk7Jun11}. As for the F1-P supertube, the product of the charge densities, $\rho^{}_1\rho^{}_2$, is constant along the curve, and this comes again from requiring that the supersymmetries be the same as those of the electric charges.

As a side note, we would like to remark that one could write down a more general Ansatz than \eqref{genPistAnsatz} by including other terms that can be constructed out of $P_i$ and $\Gamma^{0\psi}$, namely $\Gamma^{0\psi}P_i$ and $P_1P_2$.  However, if we require that we generate only one type of charge other than the original electric charges (corresponding to $P_i$) and momentum along $\psi$ (corresponding to $\Gamma^{0\psi}$), then \eqref{genPist} is the most general projector.  Another interesting fact is that the supertube projector in equation  \eqref{genPistAnsatz} can be written as:
\begin{align}
 \Pi_{st} &= U^{-1} \Pi_0 U,\qquad
 \Pi_0={1\over 2}(1+\Gamma^{0\psi}),\qquad
 U=\exp\left(
 {\theta_1\over 4}[P_1,\Gamma^{0\psi}]+{\theta_2\over 4}[P_2,\Gamma^{0\psi}]
 \right)\,,
 \label{niceRewriting} 
\end{align}
and hence changing the angle $\alpha$ can be thought as giving rise to a rotation in the
spinor space. 

\subsection{Double bubbling}
\label{apsc:doubleBubbling}

Here we consider two successive supertube transitions in which a system of three electric charges first undergoes a supertube transition and produces two dipole charges, which, in turn, undergo a supertube transition and produce another dipole charge.  We then use the methods developed above to derive the corresponding supersymmetry projectors.

\subsubsection{First supertube transition}

Consider the supertube transition of the following three electric charges:
\begin{align}
 \begin{pmatrix}D1(z)\\ D5(1234z) \\ P(z)\end{pmatrix}.
\label{D1D5Porig-app}
\end{align}
The supersymmetry projectors for this configuration  are:
\begin{align}
 \Pi_1&={1\over 2}(1+\Gamma^{0z}\sigma_1),\qquad
 \Pi_2={1\over 2}(1+\Gamma^{01234z}\sigma_1),\qquad
 \Pi_3={1\over 2}(1+\Gamma^{0z})\,.\label{fhiu4Aug10}
\end{align}
Each of these projectors preserve $16$ supersymmetries, they commute with one another  and together they leave four common supersymmetries unbroken.
Later it will be useful to consider a fourth projector:
\begin{align}
 \Pi_4&={1\over 2}(1+\Gamma^{01234z})\,.
 \label{redundproj}
\end{align}
If $\Pi_{1,2,3}\cQ=0$, then it follows that $\Pi_4\cQ=0$.  The
projector, $\Pi_4$, corresponds to KKM($1234z;\theta$), where $\theta$
is an arbitrary direction in the transverse space.  Even if we added
such a KKM to the D1-D5-P system it would not break further
supersymmetry, but we focus on the case with the three original electric
charges \eqref{D1D5Porig-app} without a fourth one.

From the supergravity analysis
\cite{Bena:2004de,Elvang:2004ds,Gauntlett:2004qy}, it is known that a
three-charge system \eqref{D1D5Porig-app} can contain a supersymmetric
black ring that has three dipole charges, corresponding to
D5($1234\theta$), D1($\theta$), and KKM($1234\theta;z$), where $\theta$
is the direction of the ring. This black ring can be thought of as a
black supertube with three dipole charges, and both the near-ring limit
\cite{Bena:2004wv} and the general solution have four supersymmetries.
However, here we consider the \emph{special} situation where there is no
KKM dipole charge, and the supertube has three charges and two dipole
charges:
\begin{align}
 \label{exbw4Aug10}
 \begin{matrix}N_1\\ N_2 \\ N_3\end{matrix}
 \begin{pmatrix}D1(z)\\ D5(1234z) \\ P(z)\end{pmatrix}
 ~~\to ~~
 \begin{matrix}n_1\\ n_2 \\  J \end{matrix}
 \begin{pmatrix}d5(1234\theta)\\ d1(\theta) \\ \cJ(\theta) \end{pmatrix}
\end{align}
where $N_i,n_i,J$ are quantized numbers of charges.  

Our first goal is to  derive the supersymmetry projector for this supertube  by using the properties of this configuration known from the DBI \cite{Bena:2004wt} and the near-tube supergravity descriptions \cite{Bena:2004wv}.  Our analysis will closely parallel  the analysis of the previous subsection, where we derived the projector for the F1-P supertube based on the known properties of the configuration.

\begin{figure}[t]
\begin{center}
 \includegraphics[height=4cm]{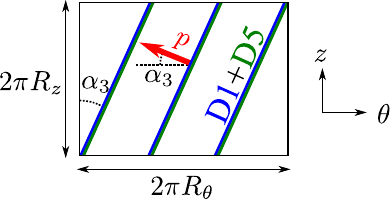} \caption{The straight
 D1-D5-P tube configuration.   }
 \label{d1d5p-tube}
\end{center}
\end{figure}

This kind of supertube transition \eqref{exbw4Aug10} has been studied by the DBI action in \cite{Bena:2004wt} and it was observed that the following relations hold among charges:
\begin{align}
 {N_1\over n_2}={N_2 \over n_1}={J\over N_3}.\label{fhtz6Aug10}
\end{align}
The same relations can also be derived by the analysis of the
supergravity black ring solution; the first relation is necessary for
the absence of closed time-like curves while the second relation follows
if we require that the configuration locally preserve one quarter of the
supersymmetry. The angular momentum $J$ and the radius $R_\theta$ of the
ring are related by \cite{Bena:2004wt}:
\begin{align}
 J={2\pi R_\theta^2}\left(n_1\tau_{D5}V_4+n_2\tau_{D1}\right),
 \label{jhdr6Aug10}
\end{align}
where $(2\pi)^4V_4$ is the 4-volume of the 1234 directions that are wrapped by the D5 branes.  This relation can be thought of as generalization of \eqref{Rpsi}.

One can do the local analysis as in Section \ref{sssapp:F1P}  by zooming in onto a very small region near a point along the round supertube, where the D1-branes and D5-branes can be thought of as straight.  However, in order not to complicate the discussion and formulae, we will simply work with a straight  supertube and keep in mind that it can be replaced by a local analysis near an arbitrary supertube.   The final formulae will be in terms of ratios of quantities and are only locally valid.

If the condition \eqref{fhtz6Aug10} is met, then the D1-branes and
D5-branes are parallel with each other in the $(\theta, z)$ plane and at
an common angle, $\alpha_3$, with the $z$ axis where
\begin{align}
 \tan\alpha_3&={n_2 R_\theta \over N_1 R_z }={n_1 R_\theta \over N_2 R_z}.\label{fzsg6Aug10}
\end{align}
The configuration of the D1 and D5 branes is shown in
Fig.~\ref{d1d5p-tube}.  Furthermore, these D1's and D5's are moving in
the direction perpendicular to their world-volume with total momentum
\begin{align}
 (p^\theta,p^z)=\left(-{J\over R_\theta},{N_3\over R_z}\right)\,.\label{nktp7Jun11}
\end{align}
So, this supertube can locally be viewed as a D1-D5 system tilted and
boosted transverse to its world-volume, which is a \nBPS{4} state; see
Fig.\ \ref{d1d5p-tube}.

To derive the projectors for this tilted-and-boosted D1-D5 system, we
follow a similar path to that of Section \ref{apsc:genericSupertube} to
derive the projector for the F1-P system.  Namely, we begin by
considering a configuration with D1($z$), D5(1234$z$), and P($\theta$)
charges; namely, the branes are aligned with the $z$ direction and have
momentum along the $\theta$ direction.  We first construct the
projectors for this configuration aligned with the $z$- and
$\theta$-axes.  Then, we will simply rotate the configuration in the
$(\theta,z)$ plane and obtain the desired projectors for the
tilted-and-boosted D1-D5 system.

As we discussed in the main text, one can view the whole process
as two separate supertube transitions of the D1-P and D5-P systems, so
that the end result is two locally-parallel and marginally bound
supertubes.
If we simply boost the D1 and D5 projectors of \eqref{fhiu4Aug10} in the
$-\theta$ direction then we trivially get the two commuting projectors:
\begin{align}
\widetilde  \Pi_{\text{D1(z)-P($\theta$)}}={1\over 2}(1+  a \, \Gamma^{0z}\sigma_1 +  b \, \Gamma^{z \theta}  )  \,,\qquad
\widetilde \Pi_{\text{D5(01234z)-P($\theta$)}}={1\over 2}(1 +  a \, \Gamma^{01234z}\sigma_1 +  b \,  \Gamma^{1234z\theta} ) \,, 
 \label{realD1D5boost}
\end{align}
where $a^2 - b^2 =1$. To convert these into the canonical forms of
projectors, as in \eqref{variousProjectors}, from which we can read off
the component charges, we multiply the first one by
$\Gamma^{0z}\sigma_1$ and the second one by $\Gamma^{01234z}\sigma_1$
and rescale by $a^{-1}$ to obtain the projectors:
\begin{align}
 \Pi_{\text{D1(z)-P($\theta$)}}={1\over 2}(1+  c_2\Gamma^{0z}\sigma_1 +  s_2\Gamma^{0\theta}  )  \,,\qquad
 \Pi_{\text{D5(01234z)-P($\theta$)}}={1\over 2}(1 +  c_2\Gamma^{01234z}\sigma_1 + s_2\Gamma^{0\theta} ) \,, 
 \label{simpD1D5boost}
\end{align}
where the angle, $\alpha_2$, is related to the boost by $c_2 =
\frac{1}{a}$, $s_2 = \frac{b}{a}$.  However, the problem with these
projectors \eqref{simpD1D5boost} is that they no longer commute with
each other, even though they are equivalent to the original set of
projectors \eqref{realD1D5boost}. Since they are equivalent to
\eqref{realD1D5boost} they do commute on their common null space as do
the projectors in \eqref{BubbProj4a} and \eqref{BubbProj4b}.

The resolution of this dilemma is to consider instead the following more
general class of projectors:
\begin{align}
\label{fhji4Aug10}
\begin{split}
  \Pi_1^{(0)}&={1\over 2}(1
 +c_1c_2\Gamma^{0z}\sigma_1
 +s_1s_2\Gamma^{01234z}\sigma_1
 +c_1s_2\Gamma^{0\theta}
 -s_1c_2\Gamma^{01234\theta}
 ),\\
 \Pi_2^{(0)}&={1\over 2}(1
 +s_1s_2\Gamma^{0z}\sigma_1
 +c_1c_2\Gamma^{01234z}\sigma_1
 -s_1c_2\Gamma^{0\theta}
 +c_1s_2\Gamma^{01234\theta}
 ),\\
 [\Pi_1^{(0)},\Pi_2^{(0)}]&=0,\qquad
 c_i=\cos\alpha_i,\qquad s_i=\sin\alpha_i,\qquad i=1,2,
\end{split}
\end{align}
where the angles, $\alpha_{1,2}$, will be determined below.  Note that,
as desired, these projectors \emph{commute} with each other.  On the
other hand, they now contain $\Gamma^{01234\theta}$ terms, which
correspond to KKM($01234\theta;z$) charges and may seem unwanted.
However, these  terms are actually acceptable if
the charges represented by $\Pi_1^{(0)}$ and the ones represented by
$\Pi_2^{(0)}$ add up to the charges that we want in the system.  Namely,
the $\Gamma^{01234\theta}$ terms cause no problem at all if the \emph{net}
KKM($01234\theta;z$) charge vanishes.

There are several reasons for considering this more general class of
projectors.  The most important one is that if one is to decompose the
boosted D1-D5 system, which preserves 8 supersymmetries, into two
sub-systems with commuting projectors, each of which preserves 16
supersymmetries, then the charges of these sub-systems are \emph{not}
those of a boosted D1 and a boosted D5, as one may naively expect. Actually, to get
\emph{commuting} projectors in the canonical form determined by
\eqref{variousProjectors}, one finds that one must include a
$\Gamma^{01234\theta}$ term which corresponds to a KKM($1234\theta;z$)
charge.  We can derive this as follows.  If we dualize the
D1($z$)-D5($1234z$)-P($\theta$) system by $T_{z\theta 12}, S, T_{\theta
13}$-dualities, we can convert it into D2(23), D2(14), D2(13).  Then,
after an $SO(2)_{12} \times SO(2)_{34}$ rotation, we can go to a frame
only with D2($1'4'$) and D2($2'3'$).  These two stacks of D2-branes are
mutually BPS and their projectors commute, and hence are the one-charge subsystems of 
our two-charge system.  If we take each sub-system and
$SO(2)_{12}\times SO(2)_{34}$ rotate it back, it now has D2(24) in
addition to D2(23), D2(14), D2(13).  If we further dualize it back to
the D1-D5-P frame, we obtain two mutually BPS sub-systems, each having
KKM($1234\theta; z$) charge in addition to the original D1($z$),
D5($1234z$) and P($\theta$) charge.  Hence, from this perspective, 
having an extra KKM($1234\theta;z$) charge is a
necessity, not an option.

Another way to arrive at the projectors above is to ask what are the
most general commuting projectors with the charges of the system.
Clearly the the projectors can contain $\Gamma^{0z}\sigma_1$, $\Gamma^{01234z}\sigma_1$
and $\Gamma^{0\theta}$ because they respectively correspond to the D1($z$),
D5($1234z$) and P($\theta$) charges. However, one can also add 
$\Gamma^{01234\theta}$ (corresponding to KKM($1234\theta;z$)), which commutes with the other charges, 
as long as the KKM charge of the total system is zero. 
Hence, our goal is to
decompose the \nBPS{4} D1($z$)-D5($1234z$)-P($\theta$) system into two
mutually BPS sub-systems each of which is \nBPS{2}, whose projectors
commute, and whose KKM charges sum up to zero. 

Yet another heuristic way of thinking of our choice of projector is that
it is the most general projector carrying these charges that does not,
at leading order in rotations, involve a direct mix of the D1 and D5
sub-systems.  The most general projector is obtained by taking four
arbitrary coefficients, $a_i$, for the gamma matrices on the right-hand
side of one of the projectors.  To be a projector one must have $\sum_i
a_i^2 =1$ and so there is a three-parameter family.  However, if one
excludes the direct rotation of the D1 system into the D5 system at
first order, and requires the two projectors to commute, one arrives at
the result in \eqref{fhji4Aug10}

We now fix the angles, $\alpha_{1,2}$, in terms of the physical
parameters.  Each of the projectors $\Pi_{i}^{(0)}$ corresponds to a
\nBPS{2} sub-system with four kinds of charges, D1($z$),
D5($1234z$), P($\theta$) and KKM($1234z;\theta$).  We
require that the total charges of the combined system satisfy:
\begin{align}
\label{chargecond_app}
\begin{split}
 M_1c_1c_2+M_2s_1s_2&=q_1,\qquad
 ~~~~M_1s_1s_2+M_2c_1c_2=q_2,\\
 M_1c_1s_2-M_2s_1c_2&=-p,\qquad
 -M_1s_1c_2+M_2c_1s_2=0.
\end{split}
\end{align}
Here $q_1=\tau_{D1}Q_z^{D1}$, $q_2=\tau_{D5}Q_z^{D5}$, and $p$ is
momentum. The masses, $M_{1,2}$, are those of the two \nBPS{2}
sub-systems that are mutually BPS\@.  The last equation in
\eqref{chargecond_app} is the one that sets the total KKM charge to
zero.  From \eqref{chargecond_app}, we can derive
\begin{align}
\begin{split}
 \cos(\alpha_1+\alpha_2)&={q_1-q_2\over M_-},\quad
 \sin(\alpha_1+\alpha_2)=-{p\over M_-},\label{jfom6Aug10}\\
 \cos(\alpha_1-\alpha_2)&={q_1+q_2\over M_+},\quad
 \sin(\alpha_1-\alpha_2)={p\over M_+},\\
 M_\pm  \equiv  M_1\pm M_2&=\sqrt{p^2+(q_1\pm q_2)^2}\,.
\end{split}
\end{align}
The physical charges, $q_1,q_2$ and $p$, are determined in terms of the
quantized charges $N_i,n_i,J$ (modulo the relation \eqref{fhtz6Aug10})
so that, after rotation in the $(\theta,z)$ plane, we end up with the
desired diagonal D1-D5-P configuration shown in Fig.\ \ref{d1d5p-tube}.
We will do that a little later. The two sub-systems have mass $M_1$ and
$M_2$ and, since the two sub-systems are mutually BPS, the mass of the
total system is simply the sum: $M_1+M_2=\sqrt{p^2+(q_1+q_2)^2}$.  This
is the correct mass of the D1($z$)-D5($z1234$) system with rest mass
$q_1+q_2$ boosted to have momentum $p$ in the $\theta$ direction.

Now we can obtain the desired projectors for the tilted-and-boosted
D1-D5 system shown in Fig.\ \ref{d1d5p-tube} by rotating the projectors
\eqref{fhji4Aug10} by an angle $-\alpha_3$ in the $(\theta, z)$ plane.
This is trivially accomplished by replacing $\Gamma^{z}$ and
$\Gamma^{\theta}$ in \eqref{fhji4Aug10} by $\Gamma^{\zh}$ and
$\Gamma^{\thetah}$ where:
\begin{equation}
\Gamma^{\zh} \equiv c_3\Gamma^{z}+s_3\Gamma^{\theta}\,,\qquad
 \Gamma^{\thetah}\equiv c_3\Gamma^{\theta}-s_3\Gamma^{z} \,. 
\label{newGammas}
\end{equation}

To summarize,  the resulting projectors, $\Pih_{i=1,2}$,  are given by
\begin{gather}
\label{jswx20Sep10}
 \Pih_{i}={1\over 2}(1+\Ph_i),
\end{gather}
with
\begin{gather}
\begin{split}
   \Ph_1=
 c_1c_2\Gamma^{0\zh}\sigma_1
 +s_1s_2\Gamma^{01234\zh}\sigma_1
 +c_1s_2\Gamma^{0\thetah}
 -s_1c_2\Gamma^{01234\thetah},
 \\
 \Ph_2=
 s_1s_2\Gamma^{0\zh}\sigma_1
 +c_1c_2\Gamma^{01234\zh}\sigma_1
 -s_1c_2\Gamma^{0\thetah}
 +c_1s_2\Gamma^{01234\thetah}.
\end{split} 
\end{gather}

We now show that the angle $\alpha_3$ is actually fixed by the dynamical
conditions \eqref{fhtz6Aug10} and \eqref{jhdr6Aug10} to be:
\begin{align}
 \alpha_3=\alpha_1-\alpha_2.\label{jfkg6Aug10}
\end{align}
From \eqref{jfom6Aug10}, we see that
\begin{align}
 \tan(\alpha_1-\alpha_2)={p\over q_1+q_2}.\label{jgvt6Aug10}
\end{align}
As mentioned before, the charges $p,q_1,q_2$ here are nothing but the P,
D1 and D5 charges in Fig.\ \ref{d1d5p-tube}.  They are computed in terms
of the charges $N_i,n_i,J$ as follows:
\begin{align}
\label{jwaa2Jun11}
\begin{split}
  p
 &= \sqrt{(p^\theta)^2+(p^z)^2}
 = \sqrt{({N_3/ R_z})^2+({J/ R_\theta})^2}
 = ({J/ R_\theta})\sqrt{1+({n_2 R_\theta/ N_1R_z})^2},\\
 q_1
 &= \tau_{D1}\sqrt{(2\pi R_z N_1)^2+(2\pi R_\theta n_2)^2}
 = 2\pi \tau_{D1}{R_z N_1} \sqrt{1+({n_2 R_\theta/ N_1 R_z})^2}\\
 q_2
 &= {\tau_{D5}V_4}\sqrt{(2\pi R_z N_2)^2+(2\pi R_\theta n_1)^2}
 = 2\pi \tau_{D5}{R_z V_4 N_2} \sqrt{1+({n_2 R_\theta/ N_1 R_z})^2}.
\end{split}
\end{align}
Here, $p$ was computed simply by \eqref{nktp7Jun11}, and $q_{1,2}$ were
computed by considering the fact that D1- and D5-branes are wrapped
diagonally with winding numbers $(N_1,n_2)$ and $(N_2,n_1)$ around a
torus of radii $R_z,R_\theta$.  Also, in the last equalities, we used
the relations \eqref{fhtz6Aug10}. If we substitute these relations
\eqref{jwaa2Jun11} into the right hand side of \eqref{jgvt6Aug10} and
use the relations \eqref{fhtz6Aug10}, \eqref{jhdr6Aug10} and \eqref{nktp7Jun11}, we obtain
$\tan\alpha_3=\tan(\alpha_1-\alpha_2)$, namely, \eqref{jfkg6Aug10}
follows.

The projectors \eqref{jswx20Sep10} with parameters $\alpha_{1,2,3}$
satisfying the relation \eqref{jfkg6Aug10} always preserve the four
common supersymmetries of the original three-charge system.  In
particular, we can write the $\Pih_i$ in terms of the fundamental set of
projectors, $\Pi_{1,\dots,3}$ of \eqref{fhiu4Aug10}, and the additional
derived projector, $\Pi_4$ of \eqref{redundproj}:
\begin{align}
 \Pih_1&=\tfrac{1}{2}[(1-\cos\phi)-\sin\phi\,\Gamma^{z\theta}]
 +(c_3-s_3\Gamma^{z\theta})(c_1 c_2\Pi_1+s_1 s_2\Pi_2)
 -(s_3+c_3\Gamma^{z\theta})(c_1 s_2\Pi_3+s_1 c_2\Pi_4),
 \notag\\
 \Pih_2&=\tfrac{1}{2}[(1-\cos\phi)-\sin\phi\,\Gamma^{z\theta}]
 +(c_3-s_3\Gamma^{z\theta})(s_1 s_2\Pi_1+c_1 c_2\Pi_2)
 +(s_3+c_3\Gamma^{z\theta})(s_1 c_2\Pi_3 -c_1 s_2\Pi_4),
 \notag
\end{align}
where $\phi\equiv\alpha_1-\alpha_2-\alpha_3$.  Note that one needs the relation \eqref{jfkg6Aug10} to show that these annihilate the supersymmetries of the original three-charge system.  We may thus use this condition of preserving the original supersymmetries as an alternative derivation of the identity  \eqref{jfkg6Aug10}.  

By using the relations \eqref{fhtz6Aug10}, \eqref{jhdr6Aug10} and
\eqref{jfom6Aug10}, we can derive the following relations for the
angles:
\begin{align}
 \label{alpha123itoM}
  \tan\alpha_3=\tan(\alpha_1-\alpha_2)
 &=\sqrt{\rho^{}_{P}\over \rho^{}_{D1}+\rho^{}_{D5}},\qquad
 \tan(\alpha_1+\alpha_2)
 =-{\sqrt{\rho^{}_{P}\left(\rho^{}_{D1}+\rho^{}_{D5}\right)}\over \rho^{}_{D1}-\rho^{}_{D5}},
\end{align}
where
\begin{align}
 \rho^{}_{D1}={\tau_{D1} N_1\over 2\pi R_\theta},\qquad
 \rho^{}_{D5}={\tau_{D5}V_4 N_2\over 2\pi R_\theta},\qquad
 \rho^{}_{P}={N_3\over (2\pi R_z)^2 R_\theta}
\end{align}
are the densities of the original electric charges D1($z$), D5($z1234$)
and P($z$), respectively, and are easy to compute just as
\eqref{kagi7Jun11}.  So, the angles $\alpha_{1,2,3}$ are determined in
terms of the ratios of the electric charge densities.  By varying these
charge densities, we can vary the angles $\alpha_{1,2,3}$ along the
curve.  However, we cannot freely vary the charge densities
$\rho^{}_{D1}$, $\rho^{}_{D5}$, $\rho^{}_{P} $.  Using the relations
\eqref{fhtz6Aug10}, it is easy to show that
\begin{align}
 \rho^{}_{D1} \rho^{}_{P}= \rho^{}_{d1} \rho^{}_{\cJ},\qquad
 \rho^{}_{D5} \rho^{}_{P}= \rho^{}_{d5} \rho^{}_{\cJ},\label{constraintrhoD1D5P}
\end{align}
where
\begin{align}
   \rho^{}_{d1} ={n_2\tau_{D1}^{}\over 2\pi R_z},\qquad
  \rho^{}_{d5} ={n_1\tau_{D5}^{}V_4\over 2\pi R_z},\qquad
 \rho^{}_{\cJ}={J\over (2\pi R_\theta)^2 R_z}
 ={n_1\tau_{D5}V_4+n_2\tau_{D1}\over 2\pi R_z}
 \label{rhod1d5J}
\end{align}
are the charge densities of D1($\theta$), D5($\theta$1234) and
P($\theta$), respectively.  Because these dipole charge densities
$\rho^{}_{d1}$ and $\rho^{}_{d5}$ are constant along the
curve, and furthermore the angular momentum density $\rho^{}_{\cJ}$ is 
equal to their sum, the charge densities $\rho^{}_{D1}, \rho^{}_{D5}, \rho^{}_{P}$
are subject to the constraint \eqref{constraintrhoD1D5P}.  Because
\eqref{constraintrhoD1D5P} imposes two conditions on three densities,
there is $3-2=1$ parameter which we can vary along the $\psi$ curve.

Much like for ordinary two-charge supertubes, the fact that the angular momentum 
density is equal to the sum of the dipole charge densities insures that the shifts to the electric Killing 
spinors brought about by the dipole charges and angular momentum cancel; this can also be seen from supergravity analysis of the near-supertube solution \cite{Bena:2004wv}.

The projectors $ \Pih_1$ and $ \Pih_2$ defined here are not the same as
those defined in \eqref{BubbProj2a}--\eqref{BubbProj3c}.  This is
because the projectors constructed here actually commute whereas those
of \eqref{BubbProj2a}--\eqref{BubbProj3c} do not commute in general
but commute on their common null space. This difference in the
commutators is directly attributable to the fact that we introduced the
polarization angle, $\alpha_1$, and arranged the D1 projector to have
KKM charge terms. Of course, to be rigorous about the second supertube
transition, one must use the commuting projectors and the procedure
outlined in the Appendix Section
\ref{Projector-derivation}. Nevertheless, as we will see below, the
projector we obtained in eq.\ (\ref{BubbProj5a}) using noncommuting
projectors is the same as the one we obtain here using the rigorous
procedure.

It is interesting to note that, just as in \eqref{niceRewriting}, the
projectors \eqref{jswx20Sep10} can be written rather concisely as
\begin{align}
\begin{split}
  \Ph_i &=U_i^{-1}\Gamma^{0\theta}U_i,\\
  U_1&=\exp\left(
 {\alpha_2-{\pi\over 2}\over 4}[P_1,\Gamma^{0\theta}]
 -{\alpha_1\over 4}[P_2,\Gamma^{0\theta}]
 +{\alpha_3\over 4}[P_3,\Gamma^{0\theta}]
 \right),\\
 U_2&=\exp\left(
 {\alpha_2\over 4}[P_1,\Gamma^{0\theta}]
 -{\alpha_1+{\pi\over 2}\over 4}[P_2,\Gamma^{0\theta}]
 +{\alpha_3\over 4}[P_3,\Gamma^{0\theta}]
 \right),\\
 P_1&=\Gamma^{0z}\sigma_1,\qquad
 P_2=\Gamma^{0z1234}\sigma_1,\qquad
 P_3=\Gamma^{0z},
\end{split}
\end{align}
and hence can also be thought of as coming from a rotation in spinor space.

\subsubsection{Second supertube transition --- double bubbling}

Locally, the supertube transition of the D1-D5-P system can be thought of as a tilted and boosted D1-D5 system and the result is a \nBPS{4} configuration.  We found commuting supersymmetry projectors, $\Pih_i$, describing this system.  Being simply a tilted and boosted version of the D1-D5 system, this system can in principle undergo a second supertube transition where it expands into new dipole charges.  The projector after such a second supertube transition is obtained simply by using the general formula
\eqref{genPist}.  The result is
\begin{align}
 \Pih ={1\over 2}( 1+s_4^2 \Ph_1+c_4^2\Ph_2-s_4c_4 \Gamma^{0\psi} +s_4c_4 \Gamma^{0\psi}\Ph_1\Ph_2 ),\label{dblbblProjector_app}
\end{align}
where $c_4=\cos\alpha_4,s_4=\sin\alpha_4$.  

The angle $\alpha_4$ is not arbitrary but is fixed in terms of charge densities by the general relation \eqref{jecc2Jun11} that any supertube transition should obey. In \eqref{jecc2Jun11}, $\rho_{1,2}$ are the densities of the electric charges associated with the commuting projectors $\Pi_{1,2}$ in \eqref{fyiv20Sep10}, and are proportional to the BPS masses for these projectors.
The BPS masses for the commuting projectors $\Pih_{1,2}$ are equal to those for $\Pih_{1,2}^{(0)}$, which are nothing but the $M_{1,2}$ defined in \eqref{jfom6Aug10}.  
Therefore, the relation \eqref{jecc2Jun11} is
\begin{align}
 \tan^2\alpha_4={M_2\over M_1}={\tan\alpha_2\over \tan\alpha_1},\label{alpha4itoM}
\end{align}
where, in the last equality,  we used \eqref{chargecond_app} and \eqref{jwaa2Jun11}.  So, the $\Pih$ defined in \eqref{dblbblProjector_app}, along with the condition \eqref{alpha4itoM}, is the projector for the second supertube transition, or double bubbling.

\subsection{The relation between projectors}
\label{apsc:theRelation}

\subsubsection{Relating the projectors}

We have carefully derived the projector \eqref{dblbblProjector_app} based on the commuting projectors $\Pih_{1,2}$.  On the other hand, in the main text, we constructed the projector \eqref{BubbProj5a} based on non-commuting projectors $\Pih_{D1,D5}$ defined in \eqref{BubbProj4a} and \eqref{BubbProj4b} and we ignored the fact that $\Pih_{D1}$  and $\Pih_{D5}$  do not commute.  The two ``double bubbled'' projectors that we have constructed  are not,  {\it a priori}, guaranteed to be the same.  However, by carefully comparing the explicit expressions of the two projectors, we can straightforwardly check that the two projectors are indeed identical with the following identification of parameters:
\begin{align}
 \alpha=\alpha_2-\alpha_1,\qquad
 \cos(2\beta)=-{\tan(\alpha_1-\alpha_2)\over \tan(\alpha_1+\alpha_2)},\qquad
 \sin(2\beta)=-{\sin(2\alpha_4)\over \cos(\alpha_1-\alpha_2)}.\label{anglesRelation}
\end{align}
Note that the angle $\alpha_4$ is related to $\alpha_{1,2}$ by \eqref{alpha4itoM}.
%

\subsubsection{How many parameters do \strata\ have ?}

For entropy-counting purposes, it is interesting to ask how much freedom one has in constructing a \stratum\, and in particular a flat \stratum. {\it A priori}, the \stratum\ projector can depend on four angles $\alpha_{1,2,3,4}$ that appear in the projector
\eqref{dblbblProjector_app}, but in order to preserve the same supersymmetries as the original branes these angles must be related via (\ref{jfkg6Aug10}) and (\ref{alpha4itoM}). 
The projector we constructed in the main text, \eqref{BubbProj5a},
also depends on two angle parameters $\alpha,\beta$, that can be related to the
angles $\alpha_{1,2,3,4}$ by \eqref{anglesRelation}. Hence it appears at first glance that the 
\stratum\ has two functions worth of degrees of freedom.

Since the four angles can in turn can be re-expressed in terms of the densities
of the D1, D5 and P original electric charges through the relations
\eqref{alpha123itoM} and \eqref{alpha4itoM}, this would imply that the three D1,D5 and P densities must satisfy one constraint. However, before the second bubbling these densities must in fact satisfy \emph{two} constraints \eqref{constraintrhoD1D5P}, which imply that D1 and the D5 densities must be proportional. It is unclear whether this proportionality relation will also apply to a \stratum. One can argue that this relation comes from requiring absence of closed time-like curves near the first-bubbled supertube profile \cite{Bena:2004wv}, and since the second bubbling has now removed the problematic region of spacetime the \stratum\ does not have to satisfy this relation. One the other hand, from the point of view of the DBI construction \cite{Bena:2004wt} this relation is an intrinsic feature of the three-charge two-dipole charge supertube that undergoes the second bubbling to become a \stratum, and hence one can argue that the \stratum\ will continue satisfying this relation. 

The final answer to this question has to await a dynamical description of a \stratum\, either via a supergravity solution or via a Born-Infeld-type analysis. The important point is that the one or two density parameters that can vary along the \stratum, together with the shape modes that give its embedding in $\cM_5$, can be functions of both $\theta$ and $\psi$. Hence the \strata\ should be parameterized by several functions of two variables, and probably have much more entropy than any other horizonless object with three charges constructed so far.



\end{document}